\documentclass[12pt]{report}%
\usepackage{ amsmath, graphics, rotating}%

\begin{document}
\large
\begin{center}{\large\bf A POSSIBLE MECHANISM OF GRAVITY}
\end{center}
\vskip 1em \begin{center} {\large Felix M. Lev} \end{center}
\vskip 1em \begin{center} {\it Artwork Conversion Software Inc.,
1201 Morningside Drive, Manhattan Beach, CA 90266, USA 
(E-mail:  felixlev@hotmail.com)} \end{center}
\vskip 1em

{\it Abstract:}
\vskip 0.5em

We consider systems of two free particles in de Sitter
invariant quantum theory and calculate the mean value
of the mass operator for such systems. It is shown that,
in addition to the well known relativistic contribution
(and de Sitter antigravity which is small when the
de Sitter radius is large), there also exists a 
contribution caused by the fact that certain 
decomposition coefficients have different 
phases. Such a contribution is negative and proportional
to the particle masses in the nonrelativistic approximation.
In particular, for a class of two-body wave functions the
mean value is described by standard Newtonian gravity
and post Newtonian corrections in General Relativity.
This poses the problem whether gravity can be explained
without using the notion of interaction at all.
We discuss a hypothesis that gravity is a manifestation 
of Galois fields in quantum physics. 

\begin{flushleft} PACS: 11.30Cp, 11.30.Ly\end{flushleft}

\begin{flushleft} Keywords: quantum theory, de Sitter invariance, gravity\end{flushleft}

\vfill\eject

\tableofcontents

\vfill\eject

\chapter{Introduction}
\label{Ch1}

In standard local quantum field theories (LQFT) the Lagrangian
of the system under consideration is usually written
as $L=L_m+L_g+L_{int}$ where $L_m$ is the Lagrangian of
'matter', $L_g$ is the Lagrangian of gauge fields and
$L_{int}$ is the interaction Lagrangian. The symmetry
conditions do not define $L_{int}$ uniquely since at
least the interaction constant is arbitrary. Nevertheless,
such an approach has turned out to be highly successful
in QED, electroweak theories and QCD. By analogy with
those theories it is usually believed that the gravitational
interaction is a consequence of the graviton exchange.

A strong indirect indication on the existence of
gravitons has been obtained in measurements
of binary pulsars (see e.g. Ref. \cite{Taylor} and the
discussion in Refs. \cite{Damour,Will,WT}), although
some physicists express reservations about such a 
conclusion (see e.g. Ref. \cite{Palle}). It is also 
worth noting that the analysis performed
in the above references is based only on classical
General Relativity (GR) while difficulties in 
constructing quantum gravity have not been 
overcome yet.
One cannot exclude a possibility that, since gravity
is the only known universal interaction, its mechanism
is not similar to that for other interactions. On the
other hand, the history of physics knows many examples 
when approaches, which seemed considerably different at
first glance, actually did not contradict each other. 

In the present paper we investigate whether standard
gravitational effects can be obtained without introducing
any interaction. Since there exist different approaches to
quantum theory in general and de Sitter invariant
quantum theory in particular, we explain our approach
in Sects. \ref{S2} and \ref{S3}. The basic facts of
de Sitter invariant quantum theories are discussed in
Chapter \ref{Ch2}. Chapter \ref{Ch3} is preparatory for
calculating the mean value of the mass operator in
Chapter \ref{Ch4}. Finally, in Chapter \ref{Ch5} we
discuss a hypothesis that gravity is a manifestation of
Galois fields in quantum physics.

\section{Motivation}
\label{S2}

The physical meaning of spacetime is one of the main problems
in modern physics. In the standard approach to elementary particle
theory it is assumed from the beginning that there exists 
a background spacetime (e.g. Minkowski or de Sitter spacetime),
and the system under consideration is described by local
quantum fields defined on that spacetime. Then by using
Lagrangian formalism and Noether theorem, one can 
(at least in principle) construct global quantized operators 
(e.g. the four-momentum operator) for the system as a whole.
It is interesting to note that after this stage has been
implemented, one can safely forget about spacetime and
concentrate his or her efforts on calculating S-matrix and 
other physical observables.   

There exist two essentially different approaches to quantum 
theory --- the standard operator approach and the path 
integral approach. We accept the operator approach. In this 
case, to be consistent, we should assume that {\it any} 
physical quantity is described by a selfadjoint operator 
in the Hilbert space of states for our system 
(we will not discuss the difference between 
selfadjoint and Hermitian operators).
Then the first question which immediately arises is that, 
even in nonrelativistic quantum mechanics, there is no 
operator corresponding to time \cite{time}. It is also 
well known that, when quantum mechanics is combined with 
relativity, there is no operator satisfying all the 
properties of the spatial position operator 
(see e.g. Ref. \cite{NW}). For these reasons the 
quantity $x$ in the Lagrangian density $L(x)$ is only 
a parameter which becomes the coordinate in the 
classical limit.

These facts were well known already in 30th of the last 
century and became very popular in 60th (recall the
famous Heisenberg S-matrix program). In the first section 
of the well-known textbook \cite{BLP} it is claimed that 
spacetime and local quantum fields are rudimentary 
notions which will disappear in the ultimate quantum theory.
Since that time, no arguments questioning those ideas have 
been given, but in view of the great success of gauge 
theories in 70th and 80th, 
such ideas became almost forgotten. 

The problem of whether the empty classical spacetime has a
physical meaning, has been discussed for a long time. In
particular, according to the famous Mach's principle,
the properties of space at a given point depend on the 
distribution of masses in the whole Universe. As described
in a wide literature (see e.g. Refs. \cite{Mach1,Mach2}
and references therein), Mach's principle was a guiding 
one for Einstein in developing GR, but when it has been 
constructed, it has been realized that it does not contain 
Mach's principle at all! As noted in Ref. \cite{Mach1,Mach2},
this problem is not closed.

At present, the predictions of the standard model are 
in agreement with experiment with an unprecedented 
accuracy. At the same time, the well-known difficulties of 
the LQFT have not been 
overcome. For this reason there exist several approaches 
(string theory, noncommutative quantum theory etc.) in 
which the product of local interacting quantum fields at 
the same spacetime point is somehow avoided. 

Consider now the problem of how 
one should define the notion of elementary particles. 

Although particles are observable and fields are not, 
in the spirit of the LQFT, fields are more fundamental 
than particles, and a possible definition is as 
follows \cite{Wein1}: 'It is simply a particle whose
field appears in the Lagrangian. It does not matter if
it's stable, unstable, heavy, light --- if its field 
appears in the Lagrangian then it's elementary, 
otherwise it's composite'.
    
Another approach has been developed by Wigner in his 
investigations of unitary irreducible representations 
(UIRs) of the Poincare group \cite{Wigner}. In view 
of this approach, one might postulate that a particle 
is elementary if the set 
of its wave functions is the space of a UIR 
of the symmetry group in the given theory 
(see also Ref. \cite{JMLL}).

Although in standard well-known theories (QED, electroweak 
theory and QCD) the above approaches are equivalent, 
the following problem arises. The symmetry 
group is usually chosen as a group of motions of 
some classical 
manifold. How does this agree with the above discussion 
that quantum theory in the operator formulation should 
not contain spacetime? A possible answer is as follows. 
One can notice that for calculating observables (e.g. the
spectrum of the Hamiltonian) we need in fact not a 
representation of the group but a representation of its 
Lie algebra by Hermitian operators. After such a 
representation has been constructed, we have only 
operators acting in the Hilbert space and this is all 
we need in the operator approach. The representation 
operators of the group are needed only if it is 
necessary to calculate some macroscopic 
transformation, e.g. time evolution. In the approximation 
when classical time is a good approximate parameter, 
one can calculate evolution, but nothing guarantees 
that this is always the case (e.g. at the very early
stage of the Universe). An interesting discussion of
this problem can be found in Ref. \cite{JMLL1}. 
Let us also note that in
the stationary formulation of scattering theory, the
S-matrix can be defined without any mentioning of
time (see e.g. Ref. \cite{Kato}). For these reasons 
we can assume that on quantum level the symmetry 
algebra is more fundamental than the symmetry group.

In other words, instead of saying that some operators 
satisfy commutation relations of a Lie algebra 
$A$ because spacetime $X$ has a group of motions $G$ such 
that $A$ is the Lie algebra of $G$, we say that there
exist operators satisfying  commutation 
relations of the Lie algebra $A$ such that: for some 
operator functions $\{O\}$ of them, the classical 
limit is a good approximation, a set $X$ of the eigenvalues
of the operators $\{O\}$ represents a classical manifold with 
the group of motions $G$ and its Lie algebra is $A$. 
This is not of course in the spirit of famous Klein's Erlangen 
program or LQFT.

Consider for illustration the well-known example of 
nonrelativistic quantum mechanics. Usually
the existence of the Galilei spacetime is assumed from
the beginning. Let $({\bf r},t)$ be the space-time coordinates
of a particle in that spacetime. Then
the particle momentum operator is $-i\partial/\partial {\bf r}$
and the Hamiltonian describes evolution by the Schroedinger equation.
In our approach one starts from an irreducible representation
of the Galilei algebra.
The momentum operator and the Hamiltonian represent four of ten
generators of such a representation. If it is implemented in 
a space of functions 
$\psi({\bf p})$ then the momentum operator is simply the 
operator of multiplication by ${\bf p}$. Then the
position operator can be {\it defined} as 
$i\partial/\partial {\bf p}$
and time can be {\it defined} as an evolution parameter such that 
evolution is described by the Schroedinger equation with
the given Hamiltonian. Mathematically the both approaches are
equivalent since they are related to each other by the Fourier
transform. However, the philosophies behind them are
essentially different. In the second approach there is 
no empty spacetime (in the spirit of Mach's principle) and 
the spacetime coordinates have a physical meaning only if
there are particles for which the coordinates can be measured.   

Summarizing our discussion, we assume that, 
{\it by definition}, on quantum level a Lie algebra is 
the symmetry algebra if there exist physical
observables such that their operators  
satisfy the commutation relations characterizing the
algebra. Then, a particle is called elementary if the 
set of its wave functions is a space of irreducible
representation of this algebra by Hermitian operators. 
Such an approach is in the spirit of that considered 
by Dirac in Ref. \cite{Dir}. 

In the literature, irreducible representations of Lie 
algebras by Hermitian operators are often called UIRs
meaning that the representation of the Lie algebra
can be extended to an UIR of the corresponding Lie group.
Although we are interested in representations of Lie
algebras, it is sometimes convenient to investigate
their properties by using well known facts about UIRs
of the corresponding Lie groups. 

\section{de Sitter invariance}
\label{S3}
 
As already mentioned, our goal is to investigate whether
the standard gravitational effects can be obtained in the
framework of a free theory. In the standard nonrelativistic
approximation gravity can be described by adding the
term $-Gm_1m_2/r$ to the nonrelativistic Hamiltonian,
where $G$ is the gravitational constant, $m_1$ and $m_2$
are the particle masses and $r$ is the distance between
the particles. Since the kinetic energy is always positive,
the free nonrelativistic Hamiltonian is positive definite
and therefore there is no way to obtain gravity in the
framework of the free theory. Analogously, in Poincare
invariant theory the spectrum of the free two-body mass
operator belongs to the interval $[m_1+m_2,\infty )$ 
while the existence of gravity necessarily requires
that the spectrum should contain values less than
$m_1+m_2$.

In theories where the invariance group is the anti
de Sitter (AdS) group SO(2,3), the structure of UIRs
of the so(2,3) algebra is well known (see e.g. Ref.
\cite{Fronsdal}). In particular, for positive
energy UIRs the AdS Hamiltonian has the spectrum in
the interval $[m,\infty )$ and $m$ has the meaning 
of the mass. Therefore the situation is pretty much
analogous to that in Poincare invariant theories.
In particular, the free two-body mass operator again
has the spectrum in the interval $[m_1+m_2,\infty )$
and therefore there is no way to reproduce gravitational
effects in the free AdS invariant theory.

Consider now the case when the de Sitter (dS) group 
SO(1,4) is chosen as the symmetry group or the dS algebra 
so(1,4) is chosen as the symmetry algebra. It is 
well known that in UIRs of the dS 
algebra, the dS Hamiltonian is not positive definite and
has the spectrum in the interval $(-\infty,+\infty)$ 
see e.g. Refs. \cite{Dobrev,Men,Moy,lev1,lev1a,lev3}). 
Note also that in contrast to the AdS
algebra so(2,3), the dS one does not have a supersymmetric
generalization. For this and other reasons it was believed 
that the dS group or algebra were not suitable for constructing 
elementary particle theory. 

In the framework of LQFT in curved spacetime (see e.g.
Ref. \cite{Narlikar} and references therein) the choice
of SO(1,4) as the symmetry group encounters 
serious difficulties which are intensively discussed in
the literature. Our approach considerably differs from
that in Refs. \cite{Narlikar}. In particular,
we do not require the existence of empty spacetime (see
the discussion in Sect. \ref{S2}). However, in Ref.
\cite{hep} we come to the same conclusion that in the 
standard approach the dS group cannot be a symmetry 
group. Nevertheless, the standard approach can be
modified as follows. Instead of requiring that UIR
should describe an elementary particle, we now
require that one UIR should describe a particle and
its antiparticle simultaneously. Then the theory 
with the dS symmetry become consistent (see Ref. \cite{hep}
for details).

It is well known that the group SO(1,4) is the symmetry 
group of the four-dimensional manifold in the five-dimensional 
space, defined by the equation 
\begin{equation}
 x_0^2-x_1^2-x_2^2-x_3^2-x_4^2=-R^2
\label{1}
\end{equation}
where a constant $R$ has the dimension of length.
The quantity $R^2$ is often written as $R^2=3/\Lambda$ where
$\Lambda$ is the cosmological constant. 
The nomenclature is such that $\Lambda < 0$ for  
the AdS symmetry while $\Lambda >0$ - for the dS one.
Elements of a map of the point $(0,0,0,0,R)$ (or $(0,0,0,0,-R)$) 
can be parametrized by 
coordinates $(x_0,x_1,x_2,x_3)$. If $R$ is very large then such a
map  proceeds to Minkowski space and the action of the dS group 
on this map --- to  the action of the Poincare group.
The recent astronomical data show that, although
$\Lambda$ is very small, it is probably positive 
(see e.g. Ref. \cite{Perlmutter}). For this reason the 
interest to dS theories has increased. Nevertheless, as noted
above, the existing difficulties have not been overcome.

\begin{sloppypar}
In the present paper it will be convenient for us to work with 
the units $\hbar/2=c=1$. Then the spin of any particle is
always an integer. For the normal relation between spin and 
statistics, the spin of fermions is odd and the spin of bosons
is even. In this system of units the representation generators 
of the SO(1,4) group
$M^{ab}$ ($a,b=0,1,2,3,4$, $M^{ab}=-M^{ba}$) should satisfy the
commutation relations
\begin{equation}
[M^{ab},M^{cd}]=-2i (\eta^{ac}M^{bd}+\eta^{bd}M^{as}-
\eta^{ad}M^{bc}-\eta^{bc}M^{ad})
\label{2}
\end{equation}
where $\eta^{ab}$ is the diagonal metric tensor such that
$\eta^{00}=-\eta^{11}=-\eta^{22}=-\eta^{33}=-\eta^{44}=1$.
\end{sloppypar}

An important observation is as follows. If we accept that the
symmetry on quantum level means that proper commutation 
relations are satisfied (see Sect. \ref{S2}) then Eq. (\ref{2})
can be treated as the {\it definition} of the dS symmetry on 
that level. In our system of units all the operators 
$M^{ab}$ are dimensionless, in contrast with the situation 
in Poincare invariant theories, where the representation 
generators of the Lorentz
group are dimensionless while the momentum operators have the
dimension $(length)^{-1}$. For this reason it is natural to
believe that the dS or AdS symmetries are more 
fundamental that the Poincare symmetry. 

Note that our definition of the dS symmetry on
quantum level does not involve the cosmological constant at 
all. It appears only if 
one is interested in interpreting results in terms of
the dS spacetime or in the Poincare limit. Since all the 
operators $M^{ab}$ are dimensionless in units $\hbar/2=c=1$,
dS (or AdS) invariant quantum theories can be formulated 
only in terms of dimensionless variables. 
In particular one might expect that the gravitational
and cosmological constants are not fundamental in the
framework of such theories. Mirmovich has proposed a 
hypothesis \cite{Mirmovich} that only
quantities having the dimension of the angular momentum
can be fundamental.

If one assumes that spacetime is fundamental then in the 
spirit of GR it is natural to think that 
the empty space is flat, i.e. that the cosmological 
constant is equal to zero. This was the subject of the 
well-known dispute between Einstein and de Sitter
described in a wide literature (see e.g. Refs. 
\cite{Mach2,Einst-dS} and references therein). In the 
modern approach to the LQFT, the cosmological constant 
is given by a contribution of vacuum diagrams, 
and the problem is to explain why it is so small. On the 
other hand, if we assume that symmetry on quantum level in 
our formulation is more fundamental, then the problem of 
the cosmological constant does not exist at all. Instead we 
have a problem of why nowadays Poincare symmetry is so 
good approximate symmetry. It seems natural to 
involve the anthropic principle for the explanation of 
this phenomenon (see e.g. Ref. \cite{Linde} and references 
therein).

\chapter{Basic properties of de Sitter invariant 
quantum theories}
\label{Ch2}

\section{UIRs of the so(1,4) algebra}
\label{S4}

There exists a wide literature devoted to 
UIRs of the dS group and algebra
(see e.g. Refs. \cite{Dix1,Tak,Hann,Str,Schwarz,
Men,Moy,Dobrev,Mielke,Klimyk,lev1,lev1a}).
In particular the first complete mathematical classification of the
UIRs has been given in Ref. \cite{Dix1}, three well known
realizations of the UIRs have been first considered in Ref. \cite{Tak}
and their physical context has been first discussed in Ref. \cite{Hann}.

It is well known that for classification of UIRs of the dS group,
one should, strictly speaking, consider not the group SO(1,4) 
itself but its
universal covering group. The investigation carried out in
Refs. \cite{Dix1,Tak,Hann,Str,Moy} has shown that this 
involves only replacement of the SO(3) group by its universal 
covering group SU(2). Since this procedure is well known then
for illustrations we will work with the SO(1,4) group itself and 
follow a very elegant presentation for physicists in terms
of induced representations, given in the book \cite{Men}
(see also Refs. \cite{Dobrev,Barut,Str}). 
The elements of the SO(1,4) group can be described in the
block form
\begin{equation}
g=\left\|\begin{array}{ccc}
g_0^0&{\bf a}^T&g_4^0\\
{\bf b}&r&{\bf c}\\
g_0^4&{\bf d}^T&g_4^4
\end{array}\right\|\ 
\label{3}
\end{equation}
where 
\begin{equation}
\label{4}
{\bf a}=\left\|\begin{array}{c}a^1\\a^2\\a^3\end{array}\right\| \quad
{\bf b}^T=\left\|\begin{array}{ccc}b_1&b_2&b_3\end{array}\right\|
\quad r\in SO(3)
\end{equation}
(the subscript $^T$ means a transposed vector).

UIRs of the SO(1,4) 
group are induced from UIRs of the subgroup $H$ defined
as follows \cite{Str,Men,Dobrev}. Each element of $H$ can be uniquely
represented as a product of elements of the subgroups
SO(3), $A$ and ${\bf T}$: $h=r\tau_A{\bf a}_{\bf T}$ where 
\begin{equation}
\tau_A=\left\|\begin{array}{ccc}
cosh(\tau)&0&sinh(\tau)\\
0&1&0\\
sinh(\tau)&0&cosh(\tau)
\end{array}\right\|\ \quad
{\bf a}_{\bf T}=\left\|\begin{array}{ccc}
1+{\bf a}^2/2&-{\bf a}^T&{\bf a}^2/2\\
-{\bf a}&1&-{\bf a}\\
-{\bf a}^2/2&{\bf a}^T&1-{\bf a}^2/2
\end{array}\right\|\ 
\label{5}
\end{equation}

The subgroup $A$ is one-dimensional and the three-dimensional
group ${\bf T}$ is the dS analog of the conventional
translation group (see e.g. Ref. \cite{Men}). We hope it 
should not cause misunderstandings when 1 is used in its
usual meaning and when to denote the unit element of the
SO(3) group. It should also be clear when $r$ is a true
element of the SO(3) group or belongs to the SO(3) subgroup
of the SO(1,4) group. 

Let $r\rightarrow \Delta(r;{\bf s})$ be a UIR of the group
SO(3) with the spin ${\bf s}$ and 
$\tau_A\rightarrow exp(i\mu\tau)$ be a
one-dimensional UIR of the group $A$, where $\mu$ is a real
parameter. Then UIRs of the group $H$ used for inducing to
the SO(1,4) group, have the form
\begin{equation}
\Delta(r\tau_A{\bf a}_{\bf T};\mu,{\bf s})=
exp(i\mu\tau)\Delta(r;{\bf s})
\label{6}
\end{equation} 
We will see below that $\mu$ has the meaning of the dS
mass and therefore UIRs of the SO(1,4) group are
defined by the mass and spin, by analogy with UIRs
in Poincare invariant theory.

Let $G$=SO(1,4) and $X=G/H$ be a factor space (or
coset space) of $G$ over $H$. The notion of the factor 
space is well known (see e.g. Refs. 
\cite{Naimark,Dobrev,Str,Men,Barut}).
Each element $x\in X$ is a class containing the
elements $x_Gh$ where $h\in H$, and $x_G\in G$ is a
representative of the class $x$. The choice of
representatives is not unique since if $x_G$ is
a representative of the class $x\in G/H$ then
$x_Gh_0$, where $h_0$ is an arbitrary element
from $H$, also is a representative of the same 
class. It is well known that $X$ can be treated 
as a left $G$ space. This means that if $x\in X$
then the action of the group $G$ on $X$ can be
defined as follows: if $g\in G$ then $gx$ is a
class containing $gx_G$ (it is easy to verify
that such an action is correctly defined). 

As noted above, although we
can use well known facts about group representations,
our final goal is the construction of the
generators. The explicit form of the generators $M^{ab}$
depends on the choice of representatives in
the space $G/H$. As explained in several
papers devoted to UIRs of the SO(1,4) group
(see e.g. Ref. \cite{Men}), to obtain
the  possible closest analogy between UIRs of
the SO(1,4) and Poincare groups, one should proceed
as follows. Let ${\bf v}_L$ be a representative 
of the Lorentz group in the factor space SO(1,3)/SO(3)
(strictly speaking, we should consider $SL(2,c)/SU(2)$).
This space can be represented as the well known velocity
hyperboloid with the Lorentz invariant measure
\begin{equation}
d\rho({\bf v})=d^3{\bf v}/v_0
\label{7}
\end{equation}
where $v_0=(1+{\bf v}^2)^{1/2}$. Let $I\in SO(1,4)$ be a
matrix which formally has the same form as
the metric tensor $\eta$. One can show 
(see e.g. Ref. \cite{Men} for details) that 
$X=G/H$ can be represented as a union of three
spaces, $X_+$, $X_-$ and $X_0$ such that 
$X_+$ contains classes ${\bf v}_Lh$, $X_-$
contains classes ${\bf v}_LIh$ and $X_0$ is of
no interest for UIRs describing elementary particles 
since it has measure zero relative to the spaces
$X_+$ and $X_-$.

As a consequence of these results, the space of UIR
of the so(1,4) algebra can be implemented as follows.  
If $s$ is the spin of the particle under 
consideration, then we
use $||...||$ to denote the norm in the space of 
UIR of the su(2) algebra with the spin $s$. 
Then the space of UIR in question is the space of 
functions $\{f_1({\bf v}),f_2({\bf v})\}$ on
two Lorentz hyperboloids with the range in the space of
UIR of the su(2) algebra with the spin $s$ and such that
\begin{equation}
\int\nolimits [||f_1({\bf v})||^2+
||f_2({\bf v})||^2]d\rho({\bf v}) <\infty
\label{8}
\end{equation}

We see that, in contrast with UIRs of the Poincare 
algebra (and AdS one), where UIRs are implemented on
one Lorentz hyperboloid, UIRs of the dS algebra can be
implemented only on two Lorentz hyperboloids, $X_+$
and $X_-$. As shown in Ref. \cite{hep}, this
fact (which is well known) has a natural explanation
if it is required that one UIR should describe a
particle and its antiparticle simultaneously.

In the case of the Poincare and AdS algebras, the positive
energy UIRs are implemented on an analog of $X_+$ and 
negative energy UIRs - on an analog of $X_-$. Since the 
Poincare and AdS groups
do not contain elements transforming these spaces
to one another, the positive and negative energy UIRs 
are fully independent. At the same time, the dS 
group contains 
such elements (e.g. $I$ \cite{Men,Dobrev,Mielke}) and for 
this reason its UIRs cannot be implemented only on 
one hyperboloid. 

\begin{sloppypar}
In Ref. \cite{hep} we have described all the technical details
needed for computing the explicit form of the generators $M^{ab}$.
In our system of units the results are as follows.
The action  of the generators on functions with the supporter in
$X_+$ has the form
\begin{eqnarray}
&&{\bf M}^{(+)}=2l({\bf v})+{\bf s},\quad {\bf N}^{(+)}==-2i v_0
\frac{\partial}{\partial {\bf v}}+\frac{{\bf s}\times {\bf v}}
{v_0+1}, \nonumber\\
&& {\bf B}^{(+)}=\mu {\bf v}+2i [\frac{\partial}{\partial {\bf v}}+
{\bf v}({\bf v}\frac{\partial}{\partial {\bf v}})+\frac{3}{2}{\bf v}]+
\frac{{\bf s}\times {\bf v}}{v_0+1},\nonumber\\
&& M_{04}^{(+)}=\mu v_0+2i v_0({\bf v}
\frac{\partial}{\partial {\bf v}}+\frac{3}{2})
\label{9}
\end{eqnarray}
where ${\bf M}=\{M^{23},M^{31},M^{12}\}$,
${\bf N}=\{M^{01},M^{02},M^{03}\}$,
${\bf B}=-\{M^{14},M^{24},M^{34}\}$, ${\bf s}$ is the spin operator,
and ${\bf l}({\bf v})=-i{\bf v}
\times \partial/\partial {\bf v}$.
At the same time, the action of the generators on 
functions with the supporter 
in $X_-$ is given by
\begin{eqnarray}
&&{\bf M}^{(-)}=2l({\bf v})+{\bf s},\quad {\bf N}^{(-)}==-2i v_0
\frac{\partial}{\partial {\bf v}}+\frac{{\bf s}\times {\bf v}}
{v_0+1}, \nonumber\\
&& {\bf B}^{(-)}=-\mu {\bf v}-2i [\frac{\partial}{\partial {\bf v}}+
{\bf v}({\bf v}\frac{\partial}{\partial {\bf v}})+\frac{3}{2}{\bf v}]-
\frac{{\bf s}\times {\bf v}}{v_0+1},\nonumber\\
&& M_{04}^{(-)}=-\mu v_0-2i v_0({\bf v}
\frac{\partial}{\partial {\bf v}}+\frac{3}{2})
\label{10}
\end{eqnarray}
\end{sloppypar}

In view of the fact that SO(1,4)=SO(4)$AT$ and $H$=SO(3)$AT$, there
also exists a choice of representatives which is probably even
more natural than that described above \cite{Men,Dobrev,Moy}. 
Namely, we can choose as
representatives the elements from the coset space SO(4)/SO(3).
Since the universal covering group for SO(4) is SU(2)$\times$SU(2)
and for SO(3) --- SU(2), we can choose as representatives the
elements of the first multiplier in the product SU(2)$\times$SU(2). 
Elements of SU(2) can be represented by the points $u=({\bf u},u_4)$
of the three-dimensional sphere $S^3$ in the four-dimensional
space as $u_4+i{\bf \sigma u}$ where ${\bf \sigma}$ are the Pauli
matrices and $u_4=\pm (1-{\bf u}^2)^{1/2}$ for the upper and
lower hemispheres, respectively. Then the calculation of the
generators is similar to that described above and the results 
are as follows.

The Hilbert space is now 
the space of functions $\varphi (u)$ on $S^3$ 
with the range in the space of the UIR of the su(2) algebra 
with the spin $s$ and such that
\begin{equation}
\int\nolimits ||\varphi(u)||^2du <\infty
\label{11}
\end{equation}
where $du$ is the SO(4) invariant volume element on $S^3$.
The explicit calculation  shows  that  the  generators for  this
realization have the form
\begin{eqnarray}
&&{\bf M}=2l({\bf u})+{\bf s},\quad {\bf B}=2\imath u_4
\frac{\partial}{\partial {\bf u}}-{\bf s}, \nonumber\\
&& {\bf N}=-2\imath [\frac{\partial}{\partial {\bf u}}-
{\bf u}({\bf u}\frac{\partial}{\partial {\bf u}})]
+(\mu +3\imath){\bf u}-{\bf u}\times {\bf s}+u_4{\bf s},\nonumber\\
&& M_{04}=(\mu +3\imath)u_4+2\imath u_4{\bf u}
\frac{\partial}{\partial {\bf u}}
\label{12}
\end{eqnarray}

Since Eqs. (\ref{8}-\ref{10}) on the one hand and
Eqs. (\ref{11}) and (\ref{12}) on  the
other  are  the  different  realization  of  one  
and   the   same
representation, there exists a unitary operator transforming
functions $f(v)$ into $\varphi (u)$ and operators 
(\ref{9},\ref{10}) into
operators (\ref{12}). For example in the spinless case the
operators (\ref{9}) and (\ref{12}) are related to each other
by a unitary transformation 
\begin{equation}
\varphi (u)=exp(-\frac{\imath}{2}\,\mu \,lnv_0)v_0^{3/2}f(v)
\label{13}
\end{equation}
where ${\bf u}={\bf v}/v_0$. 

\section{Poincare limit}
\label{S5}

A general notion of contraction has been developed in 
Ref. \cite{IW}. In our case it can be performed
as follows. Let us assume that $\mu > 0$ and denote
$m=\mu /2R$, ${\bf P}={\bf B}/2R$ and $E=M_{04}/2R$.
Then, as follows from  Eq. (\ref{9}), in the limit
when $R\rightarrow \infty$, $\mu\rightarrow \infty$
but $\mu /R$ is finite,   
one obtains a standard representation of the
Poincare algebra for a particle with the mass $m$ such 
that ${\bf P}=m{\bf v}$ is the particle momentum
and $E=mv_0$ is the particle energy. In that case
the generators of the Lorentz algebra have the same form
for the Poincare and dS algebras. Analogously the
operators given by Eq. (\ref{10}) are contracted to
ones describing negative energy UIRs of the Poincare
algebra.

In the standard interpretation of UIRs
it is assumed that each element of the full 
representation space represents a possible physical
state for the elementary particle in question. 
It is also well known (see e.g. Ref. 
\cite{Dobrev,Men,Moy,Mielke})
that the dS group contains elements (e.g. $I$)
such that the corresponding representation operator
transforms positive energy states to negative energy
ones and {\it vice versa}. Are these properties 
compatible with the fact that in the Poincare
limit there exist states with negative energies?

One might say that the choice of the energy sign is
only a matter of convention. For example, in the
standard theory we can define the energy not as
$(m^2+{\bf p}^2)^{1/2}$ but as 
$-(m^2+{\bf p}^2)^{1/2}$. However, let us consider,
for example, a system of two free noninteracting
particles. The fact that they do not interact means
mathematically that the representation describing
the system is the tensor product of single-particle
UIRs. The generators of the tensor product are equal
to sums of the corresponding single-particle
generators. In the Poincare limit the energy and
momentum can be chosen diagonal. If we assume that
both positive and negative energies are possible 
then a system of two free particles with the equal
masses can have the same quantum numbers as the
vacuum (for example, if the first particle has the
energy $E$ and momentum ${\bf p}$ while the second
one has the energy $-E$ and the momentum $-{\bf p}$)
what obviously contradicts experiment. For this
and other reasons it is well known that in the
Poincare invariant theory all the particles in
question should have the same energy sign.

We conclude that UIRs of the dS algebra cannot be 
interpreted in the standard way since such an
interpretation is physically meaningless even in
the Poincare limit. Although our approach 
considerably differs from that in LQFT in curved
spacetime, this conclusion is in agreement with
that in Refs. \cite{Narlikar} and references 
therein (see Sect. \ref{S2}). 

In the framework of our assumption that one UIR
should describe a particle and its antiparticle
simultaneously, one could try to
interpret the operators (\ref{10}) as those describing
a particle while the operators (\ref{11}) as those
describing the corresponding antiparticle. Such a
program has been implemented in Ref. \cite{hep}.
If one requires that the dS Hamiltonian should be
positive definite in the Poincare limit, then, as shown in Ref.
\cite{hep}, the annihilation and creation operators for
the particle and antiparticle in question can satisfy
only anticommutation relations, i.e. the particle and
antiparticle can be only fermions. If the normal 
spin-statistics connection takes place then it follows
that elementary particles can have only a half-integer
spin (in conventional units).  

At present the phenomenon of gravity has been observed
only on macroscopic level, i.e. for particles which
cannot be treated as elementary. Then the question
arises whether they can be described by using the
results on UIRs. The usual assumption is as follows.
In the approximation when it is possible to neglect
the internal structure of the particles (e.g. when
the distance between them is much greater than their
sizes), the structure of the internal wave function
is not important and one can consider only the part
of the wave function describing the motion of the
particle as the whole. This part is described by the
same parameters as the wave function of the elementary
particle. Even if only fermions can be elementary,
this does not mean of course that the external wave
function of the composite system can be used only if
the total spin is half-integer (in conventional
units). Moreover, the rotation of a macroscopic system 
as a whole is usually the effect which does not play
an important role in the gravitational interaction.
For this reason it is usually sufficient to describe
the motion of the macroscopic system as a whole by
using wave functions of UIRs with zero spin.  

\section{de Sitter antigravity}
\label{S6}

Consider now the Poincare limit in the approximation
when $R$ is large but the first order corrections 
in $1/R$ to the conventional energy and momentum are
taken into account. By using the definitions of the
Poincare mass, energy and momentum from the 
preceding section and taking the dS generators in
the form (\ref{9}), one can obtain the expressions
for the conventional energy and momentum in first
order in $1/R$. For simplicity we assume that the
particles is spinless and nonrelativistic. Consider
a system of two free particles with the masses 
$m_1$ and $m_2$. Then the momentum and energy
operators for each particle are given by 
\begin{eqnarray}
&& {\bf P}_j={\bf p}_j+\frac{im_j}{R} \frac{\partial}
{\partial {\bf p}_j}\nonumber\\
&& E_j=m_j + \frac{{\bf p}_j^2}{2m_j} +\frac{i}{R}({\bf p}_j
\frac{\partial}{\partial {\bf p}}_j+\frac{3}{2})
\end{eqnarray}
\label{AG1}
where ${\bf p}_j=m{\bf v}_j$ and $j=1,2$.

As noted in the preceding section, the fact that the
particle do not interact with each other implies that
the generators for the two-body system are equal
to sums of the corresponding single-particle
generators. Adding the corresponding operators and
introducing the standard total and relative momenta
\begin{equation}
{\bf P}={\bf p}_1+{\bf p}_2\quad 
{\bf q} = (m_2{\bf p}_1-m_1{\bf p}_2)/(m_1+m_2)
\label{AG2}
\end{equation}
one can obtain the expressions for the momentum ${\bf P}$ 
and energy $E$ of the two-body system as a whole. Let $M$ be
the mass operator of the two-body system defined as
$M^2=E^2-{\bf P}^2$. Then a simple calculation shows
that in our approximation
\begin{equation}
M = m_1+m_2 +\frac{{\bf q}^2}{2m_{12}}+
\frac{i}{R}({\bf q}\frac{\partial}{\partial 
{\bf q}}+\frac{3}{2})
\label{AG3}
\end{equation}
where $m_{12}=m_1m_2/(m_1+m_2)$ is the reduced mass.

In spherical coordinates the nonrelativistic mass
operator can be written as
\begin{equation}
M_{nr}=\frac{q^2}{2m_{12}} + V,\quad 
V=\frac{i}{R}(q\frac{\partial}{\partial q}+\frac{3}{2})
\label{AG4}
\end{equation}
where $q=|{\bf q}|$.
Although this expression has been obtained in first
order in $1/R$, let us consider for illustrative
purposes the spectrum of this operator. 
It acts in the space of functions $\psi(q)$ such that
\begin{equation}
\int_{0}^{\infty}|\psi(q)|^2q^2dq <\infty
\label{AG5}
\end{equation}
and the eigenfunction $\psi_K$ of $M_{nr}$ with the
eigenvalue $K$ satisfies the equation
\begin{equation}
q\frac{d\psi_K}{dq}=\frac{iRq^2}{2m_{12}}\psi_K-
(\frac{3}{2}+iRK)\psi_K
\label{AG6}
\end{equation}
The solution of this equation is
\begin{equation}
\psi_K=\sqrt{\frac{R}{2\pi}}q^{-3/2}
exp(\frac{iRq^2}{4m_{12}}-iRKlnq)
\label{AG7}
\end{equation}
and the normalization condition is
\begin{equation}
(\psi_K,\psi_{K'})=\delta(K-K')
\label{AG8}
\end{equation}

The spectrum of the operator $M_{nr}$ obviously
belongs to the interval $(-\infty,\infty)$ and one
might think that this is unacceptable. Suppose
however that $f(q)$ is a wave function of some
state. As follows from Eq. (\ref{AG7}), the probability 
to have the value of the kinetic energy $K$ in this 
state is given by
\begin{equation}
c_K=\sqrt{\frac{R}{2\pi}}\int_{0}^{\infty}
exp(-\frac{iRq^2}{4m_{12}}+iRKlnq)f(q)\sqrt{q}dq
\label{AG9}
\end{equation}
If $f(q)$ does not contain a rapidly oscilating
exponent depending on $R$ (in particular when it
does not depend on $R$) and $R$ is very large 
then $c_K$ will practically be
different from zero only if the integrand in Eq. 
(\ref{AG9}) has a stationary point $q_0$. It is
obvious that the stationary point is defined by
the condition $K=q_0^2/2m_{12}$. Therefore, for
negative $K$, when the stationary point is absent,
the value of $c_K$ will be very small.

We see that if one works only with a subset of 
wave functions not depending on $R$ (which is 
typically the case), then the existence of the
points of the spectrum of the two-body mass
operator with the values less than $m_1+m_2$
does not play an important role. 

\section{de Sitter antigravity in quasiclassical
approximation}
\label{classical}

In conventional quantum mechanics the motion of a
particle is quasiclassical if at each moment of time
$t=t_0$ the particle wave function satisfies the
following conditions (see e.g. Ref. \cite{LL}). In the
coordinate representation the function has a sharp
maximum at some ${\bf r}={\bf r}_0$, and the uncertainty
of the position $\Delta {\bf r}$ is much less than 
${\bf r}_0$. At the same time, in the velocity 
representation it should have a sharp maximum at some 
${\bf v}={\bf v}_0$, and the uncertainty
of the velocity $\Delta {\bf v}$ should be much less than 
${\bf v}_0$. In particular, the particle cannot be 
quasiclassical if it is at rest, i.e. ${\bf v}_0=0$.

As follows from this definition, the notion of
quasiclassical approximation necessarily implies that
the position and velocity operators are well-defined
and have a clear physical meaning. This is indeed the
case in conventional nonrelativistic quantum mechanics.
It is well known (see also Sect. \ref{S2}) that in 
relativistic quantum theory there is no operator
satisfying all the requirements for the position 
operator \cite{NW}. In dS invariant theories there
exists an analogous problem. In particular, as seen
from Eq. (\ref{9}), the operator ${\bf v}$ by itself
does not define the dS momentum (which is a physical
operator) uniquely, and the operator 
$i\partial /(m\partial {\bf v})$, which in nonrelativistic
quantum mechanics is the position operator in velocity
representation, does not define the physical Lorentz boost
operators uniquely. However, as noted in Sect. \ref{S5},
when $R$ is very large, the generators (\ref{9}) can
be contracted to standard generators of the UIR of the
Poincare group. In this case the momentum ${\bf P}$ is 
exactly proportional to ${\bf v}$ and the proportionality
coefficient is the mass. Moreover, when the particle is
nonrelativistic, then, as follows from Eq. (\ref{9}),
the Lorentz boost operators are proportional to the
corresponding coordinate operators in velocity 
representation. 

We conclude that, at least when $R$ is large and
$|{\bf v}|\ll 1$, there exists a well-defined
quasiclassical approximation in the representation
when the generators are given by Eq. (\ref{9}). 
For example, the wave function can be chosen in the
form 
\begin{equation}
f({\bf v})=a({\bf v})exp(-im{\bf v}{\bf r}_0) 
\label{49}
\end{equation}
where $a({\bf v})$ has a sharp maximum at ${\bf v}={\bf v}_0$
with a width $|\Delta {\bf v}|\ll |{\bf v}_0|$ and such
that $|\partial a({\bf v})/\partial {\bf v}|\ll m|{\bf r}_0|$.
A possible choice of $a({\bf v})$ is 
\begin{equation}
a({\bf v})=cv_0^{1/2}exp[-\frac{b^2}{2}({\bf v}-{\bf v}_0)^2]
\label{50}
\end{equation}
where the function is normalized to one 
(see Eq. (\ref{8})) if 
$$c=\sqrt{2}b^{3/2}/\pi^{3/4}.$$ 
Then the condition 
$|\Delta {\bf v}|\ll |{\bf v}_0|$ is satisfied if
$b|{\bf v}_0|\gg 1$ and the condition 
$|\Delta {\bf r}|\ll |{\bf r}_0|$ is satisfied if 
$b\ll m|{\bf r}_0|$. For macroscopic particles there 
exists a wide range of values $b$ such that these
conditions can be satisfied.

On classical level the effect of the additional
term in Eq. (\ref{AG3}) in comparison with the
standard free nonrelativistic expression can be
investigated as follows. We define the position
operator ${\bf r}$ as 
${\bf r}=i(\partial /\partial {\bf q} )$.
Then the classical Hamiltonian of the internal
motion corresponding to the operator (\ref{AG3}) is
\begin{equation}
H({\bf r}, {\bf q}) =\frac{{\bf q}^2}{2m_{12}} +
\frac{{\bf r}{\bf q}}{R}
\label{AG10}
\end{equation}
From classical equations of motion it follows that
$d^2{\bf r}/dt^2={\bf r}/R^2$. It is well known that
in classical dS space there exists a universal 
repulsion (antigravity) the force is which is
proportional to the distance between particles.
Therefore the operator $V$ indeed corresponds to
the dS antigravity. 

Although the example of the dS antigravity considered
in the preceding and this sections is extremely simple, 
we can draw the following three very important conclusions.

The first conclusion is that the standard classical
dS antigravity has been obtained from a quantum
operator without introducing any classical background.
When the position operator is defined as 
${\bf r}=i(\partial /\partial {\bf q} )$ and time is
defined by the condition that the Hamiltonian is the
evolution operator then we recover the classical
result obtained by considering a motion of particles
in the classical dS spacetime. This is an illustration
of the discussion in Sect. \ref{S2} about the difference
between the standard approach, where the classical dS
spacetime is introduced from the beginning, and our one.

The second conclusion is as follows. We have considered
the particles as free, i.e. no interaction into the
two-body system has been introduced. However, we have
realized that when the two-body system in the dS 
invariant theory is considered from the point of view of 
the Galilei invariant theory, the particles interact 
with each other. Although the reason of the effective 
interaction in our example is obvious, the existence of
the dS antigravity poses the problem whether other 
interactions, e.g. gravity, can be treated as a result of 
transition from a higher symmetry to Poincare or Galilei 
one.

Finally, the third conclusion is that if in a free
theory the spectrum of the mass operator has values less
than $m_1+m_2$, this does not necessarily mean that the 
theory is unphysical.

\section{Free and interacting systems}
\label{S7} 

The above discussion poses the following problem.
We treated a system of two particles as free if the
two-body generators are sums of the corresponding
single-particle generators, and in the preceding 
sections we assumed
that the single-particle generators are given by Eq. 
(\ref{9}). 
In Sect. \ref{S4} we have shown that there also exists
a realization of the single-particle generators in the
form of Eq. (\ref{12}) and in general there exist
infinitely many unitarily equivalent realizations of
single-particle generators. We have noted that in the
spinless case the realizations (\ref{9}) and (\ref{12})
are related to each other by a unitary transformation
(\ref{13}). Suppose now that we treat a two-body system
as free if the two-body generators are sums of the
single-particle generators in the form (\ref{12}), not
(\ref{9}). Then it is easy to see that in terms of the
generators (\ref{9}) the system is not free but the
sums of the two-body generators (\ref{9}) are
unitarily equivalent to the corresponding sums of
the two-body generators (\ref{12}). Therefore the
problem arises whether one should treat a system 
as free or interacting when the generators are unitarily 
equivalent to free ones.

In Poincare invariant theory a system is treated as
free or interacting depending on whether the S matrix
is the identity operator or not. For a pair of
Hamiltonians $(H,H_0)$ the S matrix is defined as
follows. First one defines the Moeller wave operators 
$W_{\pm}$ as
\begin{equation}
W_{\pm}=s-lim\,\, exp(iHt)exp(-iH_0t)P_0
\label{int1}
\end{equation}
where $t\rightarrow \pm \infty$, $s-lim$ means the 
strong limit and $P_0$ is the projector onto the subspace
corresponding to the (absolutely) continuous spectrum of
the operator $H_0$. (see e.g. Ref. \cite{Kato}). Then 
the S matrix is defined as $S=W_+^*W_-$. The Moeller 
operators are not necessarily unitary but the S matrix is.

The usual assumption is that the operator $H_0$, which is
treated as free, has only the continuous spectrum and
therefore $P_0=1$. The interacting Hamiltonian $H$ in
general can have both, the continuous and discrete
spectrum. It is well known that two selfadjoint operators
can be unitarily equivalent if and only if they have the
same spectrum (see e.g. Ref. \cite{Kato}). Therefore, if
$H$ has the discrete spectrum, the operators $H$ and
$H_0$ cannot be unitarily equivalent. In that case the
Moeller operators are not unitary. On the contrary, if
$H$ has only continuous spectrum then a typical situation
is that the spectra of $H$ and $H_0$ are the same. In
that case the Moeller operators are unitary and
\begin{equation}
H=W_{\pm}H_0W_{\pm}^{-1}
\label{int2}
\end{equation}
but nevertheless in the general case the S matrix is not the
identity operator. Therefore in general, the case when the 
generators are unitarily equivalent to the free ones should 
be treated as interacting. 

The philosophy of quantum mechanics is such that if one 
has a set of states and operators acting on these states
then only probabilities of different experiment outcomes
are important. If one applies a unitary transformation to
the given set of states and transforms the operators
accordingly then the probabilities remain the same.
From this point of view it is not quite clear why in the
case when $H$ and $H_0$ are unitarily equivalent they 
nevertheless are not on equal footing (one of them is treated
as free and the other as interacting). In the
full Hilbert space we can always find states on which $H_0$
acts as $H$ and {\it vice versa}. It is often assumed 
that the particles are free not only if the two-body
generators are the sums of the corresponding single-particle
generators but also when the wave function of the system as
a whole is a product of single-particle wave functions. 
However, the Hilbert space for the two-body system is the 
tensor product of the single-particle Hilbert spaces,
and therefore any linear combination of the products of
the single-particle wave functions is an allowable wave
function. The example considered at the end of this section
will shed some light on this problem.
In any case, if $H$ and $H_0$ are unitarily equivalent then,
the interaction can be introduced not only as $H=H_0+V$
but also by using the unitary operator realizing the
equivalence. For example, one can use the Moeller operators 
which relate 
the free and interacting Hamiltonians according to Eq. 
(\ref{int2}). Such an approach has been proposed by
Sokolov and Shatny in Ref. \cite{Sokolov,SokSh}.

Let us also note that in Poincare invariant theory there
is no unique way of introducing interaction into the
$N$-body system. Dirac has proposed a notion of forms
of relativistic dynamics and singled out three forms:
instant, front and point ones \cite{Dir}. As shown by
Sokolov and Shatny \cite{SokSh}, all the three forms 
have unitarily
equivalent S matrix and therefore they are physically
equivalent.

While in Poincare invariant theory the free operators
are defined in fact uniquely (up to unitary 
transformations of
particle spin variables), in dS invariant theory 
it is not clear what choice of operators is most
justified and there is no S matrix. As already
noted, the choice (\ref{9}) corresponds to our
intuition in Poincare invariant theory but this
does not mean that such a choice is more physical 
than say the choice (\ref{12}). We will argue below
that the choice (\ref{12}) is in fact more physical.

In any case, dS invariant theories have a feature
which is specific only for such theories, and the
reason is as follows (see Ref. \cite{lev1a} for
details). Although the dS antigravity is typically
small when $R$ is large, as follows from Eq. (\ref{AG4}),
the universal repulsion at asymptotically large
distances is dominant in comparison with any
reasonable interaction. Therefore there is no bound
states in the dS invariant theory: any state which
in Poincare invariant theory is bound becomes 
quasibound with a large lifetime. As a result, the
free and interacting operators have the same spectrum
and therefore they are unitarily equivalent. This
again poses the problem about the meaning of interactions
in dS invariant theories.

For illustration, consider a radial motion of two 
quasiclassical particles with the relative momentum 
$q_0$ and such that the distance between them is 
$r_0$. This implies that the wave function in
$q$ representation has a sharp maximum at $q=q_0$,
and the wave function in $r$ representation has a
sharp maximum at $r=r_0$. The sharpness of the
maximum in $q$ representation is defined by the
quantity $(\Delta q)/q \ll 1$ where $\Delta q$ is the
width of the maximum. Analogously the sharpness 
of the maximum in $r$ representation is defined by the
quantity $(\Delta r)/r \ll 1$. Let us take the radial wave
function for such a system and multiply it by  
the factor
\begin{equation}
\eta(q,r_0)=exp(iRGm_1m_2lnq/r_0)
\label{factor}
\end{equation}
Then, instead of Eq. (\ref{AG9}), the mass distribution
is defined by 
\begin{equation}
c_K=\sqrt{\frac{R}{2\pi}}\int_{0}^{\infty}
exp(\frac{i}{r_0}RGm_1m_2lnq-\frac{iRq^2}{4m_{12}}+
iRKlnq)f(q)\sqrt{q}dq
\label{exp1}
\end{equation}
The exponent index has the stationary point at 
$q$ satisfying the condition
\begin{equation}
K=K(q) =\frac{q^2}{2m_{12}}-\frac{Gm_1m_2}{r_0}
\label{exp2}
\end{equation}
and hence the mass distribution has a sharp
maximum about $K=K(q_0)$.

Therefore, although no operator corresponding to
gravitational interaction has been introduced, for some
class of wave functions the mean value of the mass
operator is 
compatible with the standard classical gravity. However,
the problem arises what is the physical meaning of
wave functions containing the factor (\ref{factor}).
Indeed, if the position operator is defined as above
then the distance between the particles will be not
$r_0$ but an anomalously large value. One can try
to modify the position operator, but the usual
approach is such that the position operators are
always the same regardless of whether an interaction
is present or not. 

For this reason we reformulate 
the problem as follows. We will not assume anymore
that the two-particle wave function contains the factor
(\ref{factor}) but instead introduce an
interaction into the two-body mass operator as follows.
Define the selfadjoint operator
\begin{equation}
A =\frac{1}{2} Gm_1m_2\{lnq,\frac{1}{r}\}
\label{exp3}
\end{equation}
where $\{...\}$ is used to denote the anticomutator,
and define the nonrelativistic mass operator as
\begin{equation}
{\hat M}=exp(-iRA)[\frac{q^2}{2m_{12}} +  
\frac{i}{R}(q\frac{\partial}{\partial q}+\frac{3}{2})]
exp(iRA)
\label{exp4}
\end{equation}
With such a mass operator, its eigen functions are given
by (compare with Eq. (\ref{AG7}))
\begin{equation}
{\hat \psi}_K=exp(-iRA)\sqrt{\frac{R}{2\pi}}q^{-3/2}
exp(\frac{iRq^2}{4m_{12}}-iRKlnq)
\label{exp5}
\end{equation}

\begin{sloppypar}
Let us now consider wave functions for which
the $r$-distribution is much sharper than the 
$q$-distribution, i.e. 
$(\Delta r/r)\ll (\Delta q/q)$ (recall that $\Delta r$ is
a characteristic of the wave function describing the
motion as a whole; for the classical particle it has 
nothing to do with the particle dimension). 
Since the uncertainty
of the quantity $r^{-1}$ is $\Delta r^{-1}=(\Delta r)/r^2$,
we have that $(\Delta r^{-1})/r^{-1}=(\Delta r)/r$.
Therefore the distribution in $r^{-1}$ is much sharper 
than the distribution in $q$. As a result, with a good
accuracy the action of $A$ on such wave functions
$\psi$ can be written as $A\psi=(Gm_1m_2lnq/r_0)\psi$ 
and the mass distribution is again given by Eq. (\ref{exp1}).
\end{sloppypar}

Although the above example is mainly illustrative, it 
gives grounds to draw the following
conclusions. Even if no interaction into the two-body
mass operator is introduced, for some
class of wave functions the mean value of the mass
operator is 
compatible with the standard classical gravity. This
result can be interpreted in a standard way by 
introducing an interaction by means of a unitary
operator as in Eq. (\ref{exp4}). On classical 
level the operator (\ref{exp4}) results in the same 
observable consequences as the operator
\begin{equation}
M_{standard}=\frac{q^2}{2m_{12}} +  
\frac{i}{R}(q\frac{\partial}{\partial q}+\frac{3}{2})-
\frac{Gm_1m_2}{r}
\label{exp6}
\end{equation}
which is usually expected to be a quantum generalization
of the conventional nonrelativistic gravity. Since 
at present gravity is known only on classical level,
one cannot conclude that the operator (\ref{exp4})
is inferior with respect to the operator (\ref{exp6}).
These operators are essentially different only on
quantum level. 

Below we consider in detail the case when
the generators are defined by Eq. (\ref{12}) and the 
two-body mass operator is not decomposed in powers of $1/R$. 

\chapter{de Sitter invariant quantum theory in 
su(2)$\times$su(2) basis}
\label{Ch3}

\section{UIRs in the su(2)$\times$su(2) basis}
\label{S8}

Proceeding from the method of su(2)$\times$su(2) shift
operators, developed by Hughes \cite{Hug} for constructing
UIRs of the group SO(5), and following Ref. \cite{lev3},
we now give a pure algebraic description of UIRs of
the so(1,4) algebra. It will be convenient for us to  deal
with the set of operators $({\bf J}',{\bf J}",R_{ij})$ ($i,j=1,2$)
instead  of $M^{ab}$.
Here ${\bf J}'$ and ${\bf J}"$ are two independent su(2) algebras
(i.e. $[{\bf J}',{\bf J}"]=0$).
In each of them one chooses as the basis the operators  
$(J_+,J_-,J_3)$ such that in our system of units 
$J_1=J_++J_-$, $J_2=-\imath (J_+-J_-)$ and the commutation
relations have the form
\begin{equation}
[J_3,J_+]=2J_+,\quad [J_3,J_-]=-2J_-,\quad [J_+,J_-]=J_3
\label{14}
\end{equation}

The commutation relations of the operators ${\bf J}'$ and
${\bf J}"$  with $R_{ij}$ have the form
\begin{eqnarray}
&&[J_3',R_{1j}]=R_{1j},\quad [J_3',R_{2j}]=-R_{2j},\quad
[J_3",R_{i1}]=R_{i1},\nonumber\\
&& [J_3",R_{i2}]=-R_{i2},\quad
[J_+',R_{2j}]=R_{1j},\quad [J_+",R_{i2}]=R_{i1},\nonumber\\
&&[J_-',R_{1j}]=R_{2j},\quad [J_-",R_{i1}]=R_{i2},\quad
[J_+',R_{1j}]=\nonumber\\
&&[J_+",R_{i1}]=[J_-',R_{2j}]=[J_-",R_{i2}]=0,\nonumber\\
\label{15}
\end{eqnarray}
and the commutation relations of the operators $R_{ij}$ 
with each other have the form
\begin{eqnarray}
&&[R_{11},R_{12}]=2J_+',\quad 
[R_{11},R_{21}]=2J_+",\nonumber\\
&& [R_{11},R_{22}]=-(J_3'+J_3"),\quad 
[R_{12},R_{21}]=J_3'-J_3"\nonumber\\
&& [R_{11},R_{22}]=-2J_-",\quad [R_{21},R_{22}]=-2J_-'
\label{16}
\end{eqnarray}

The relation between the sets $({\bf J}',{\bf J}",R_{ij})$ and
$M^{ab}$  is given by
\begin{eqnarray}
&&{\bf M}={\bf J}'+{\bf J}", \quad {\bf B}={\bf J}'-{\bf J}",
\quad M_{01}=\imath (R_{11}-R_{22}), \nonumber\\
&& M_{02}=R_{11}+R_{22}, \quad  
M_{03}=-i(R_{12}+R_{21}),\nonumber\\
&&  M_{04}=R_{12}-R_{21}
\label{17}
\end{eqnarray}
Then it is easy to see that Eq. (\ref{2}) 
follows from Eqs. (\ref{15}-\ref{17}) and {\it vice versa}.

Consider the space of maximal  $su(2)\times su(2)$  vectors,
i.e.  such vectors $x$ that $J_+'x=J_+"x=0$. Then from
Eqs. (\ref{15}) and (\ref{16}) it follows that the operators
\begin{eqnarray}
&&A^{++}=R_{11}\quad  A^{+-}=R_{12}(J_3"+1)-
J_-"R_{11},  \nonumber\\
&&A^{-+}=R_{21}(J_3'+1)-J_-'R_{11}\nonumber\\
&&A^{--}=-R_{22}(J_3'+1)(J_3"+1)+J_-"R_{21}(J_3'+1)+\nonumber\\
&&J_-'R_{12}(J_3"+1)-J_-'J_-"R_{11}
\label{18}
\end{eqnarray}
act invariantly on this space.
The notations are related to the property  that
if $x^{kl}$  ($k,l>0$) is the maximal su(2)$\times$su(2)
vector and simultaneously
the eigenvector of operators $J_3'$ and $J_3"$ with the 
eigenvalues $k$ and $l$, respectively, then  $A^{++}x^{kl}$
is  the  eigenvector  of  the  same
operators with the values $k+1$ and $l+1$, $A^{+-}x^{kl}$ - the
eigenvector  with
the values $k+1$ and $l-1$, $A^{-+}x^{kl}$ - 
the eigenvector with the values  $k-1$ and $l+1$ and 
$A^{--}x^{kl}$ - the eigenvector with the
values $k-1$ and $l-1$.

As follows from Eq. (\ref{14}), the vector $x_{ij}^{kl}
=(J_-')^i(J_-")^jx^{kl}$ is
the eigenvector of the operators $J_3'$ and $J_3"$ with 
the eigenvalues $k-2i$ and $l-2j$, respectively. 
Since 
$${\bf J}^2=J_3^2-2J_3+4J_+J_-=J_3^2+2J_3+4J_-J_+$$
is the Casimir operator for the ${\bf J}$  algebra, and the
Hermiticity condition can be written as $J_-^*=J_+$, 
it  follows  in  addition that 
\begin{equation}
{\bf J}^{'2}x_{ij}^{kl}=k(k+2)x_{ij}^{kl},\quad
{\bf J}^{"2}x_{ij}^{kl}=l(l+2)x_{ij}^{kl}
\label{19}
\end{equation}
\begin{equation}
J_+'x_{ij}^{kl}=i (k+1-i)x_{i -1,j}^{kl},
\quad  J_+"x_{ij}^{kl}=j (l+1-j)x_{i,j -1}^{kl}
\label{20}
\end{equation}
\begin{equation}
(x_{ij}^{kl},x_{ij}^{kl}))=
\frac{i !j !k!l!}{(k-i !)(l-j !)}(x^{kl},x^{kl})
\label{21}
\end{equation}
where $(...,...)$ is the scalar product in the representation space.
From these formulas it follows that the action of 
the operators ${\bf J}'$
and ${\bf J}"$ on $x^{kl}$  generates a space with the  
dimension $(k+1)(l+1)$  and  the
basis $x_{ij}^{kl}$ ($i=0,1,...k$, $j=0,1,...l$).
Note that the vectors $x_{ij}^{kl}$   are
orthogonal but in this section we do not normalize them to 
one (the reason will be clear below).

The Casimir operator of the second order for the 
algebra  (\ref{2}) can be written as
\begin{eqnarray}
&&I_2 =-\frac{1}{2}\sum_{ab} M_{ab}M^{ab}=\nonumber\\
&&4(R_{22}R_{11}-R_{21}R_{12}-J_3')-
2({\bf J}^{'2}+{\bf J}^{"2})
\label{22}
\end{eqnarray}
A direct calculation shows that for the
generators given by Eqs. (\ref{9}), (\ref{10}) and 
(\ref{12}), $I_2$  has the numerical  value
\begin{equation}
I_2 =w-s(s+2)+9
\label{23}
\end{equation}
where $w=\mu^2$. As noted in Sect. \ref{S5},
$\mu = 2mR$ where $m$ is the conventional mass. If $m \neq 0$
then $\mu$ is very large since $R$ is very large. We conclude
that for massive UIRs the quantity $I_2$ is a large positive
number.

 The basis in the representation space
can be explicitly constructed assuming that there exists a
vector $e^0$ which is the maximal su(2)$\times$su(2)
vector such that
\begin{equation}
J_3'e_0=n_1e_0\quad J_3"e_0=n_2e_0
\label{24}
\end{equation}
and $n_1$ is the minimum possible eigenvalue of $J_3'$ in
the space of the maximal vectors. Then $e_0$ should also
satisfy the conditions
\begin{equation}
A^{--}e_0=A^{-+}e_0=0
\label{25}
\end{equation}
We use ${\tilde I}$ to denote the operator 
$R_{22}R_{11}-R_{21}R_{12}$.
Then as follows from Eqs. (\ref{15}), (\ref{16}), (\ref{18}),
(\ref{22}), (\ref{24}) and (\ref{25}),
$${\tilde I}n_1e_0=2n_1(n_1+1)e_0.$$
Therefore, if $n_1\neq 0$ the vector $e_0$ is the eigenvector
of the operator ${\tilde I}$ with the eigenvalue 
$2(n_1+1)$ and the
eigenvector of the operator $I_2$ with the eigenvalue
$$-2[(n_1+2)(n_2-2)+n_2(n_2+2)].$$ 
The latter is obviously incompatible with Eq. (\ref{23})
for massive UIRs. Therefore the compatibility can be
achieved only if $n_1=0$. In that case we use $s$ to denote
$n_2$ since it will be clear soon that the value of $n_2$
indeed has the meaning of spin. Then, as follows from 
Eqs. (\ref{23}) and (\ref{24}), the vector $e_0$ should 
satisfy the conditions  
\begin{eqnarray}
&&{\bf J}'e^0=J_+"e^0=0,\quad J_3"e^0 =se^0, \nonumber\\
&&I_2e^0 =[w-s(s+2)+9] e^0
\label{26}
\end{eqnarray}
where $w,s>0$ and $s$ is an integer.  

Define the vectors
\begin{equation}
e^{nr}=(A^{++})^n(A^{+-})^re^0
\label{27}
\end{equation}
Then a direct calculation taking into account Eqs.
(\ref{14})-(\ref{16}), (\ref{18}), (\ref{19}), (\ref{22}), 
(\ref{25}) and (\ref{26}) gives
\begin{equation}
A^{--}A^{++}e^{nr}=-\frac{1}{4}(n+1)(n+s+2)[w+(2n+s+3)^2]e^{nr}
\label{28}
\end{equation}
\begin{equation}
A^{-+}A^{+-}e^{nr}=-\frac{1}{4}(r+1)(s-r)[w+1+(2r-s)
(2r+2-s)]e^{nr}
\label{29}
\end{equation}
\begin{equation}
(e^{n+1,r},e^{n+1,r})=\frac{(n+1)(n+s+2)[w+(2n+s+3)^2]}
{4(n+r+2)(s-r+n+2)}(e^{nr},e^{nr})
\label{28a}
\end{equation}
\begin{eqnarray}
&&(e^{n,r+1},e^{n,r+1})=\frac{1}{4}(r+1)(s-r)[w+1+(2r-s)
(2r+2-s)]\nonumber\\
&&\frac{s-r+n+1}{s-r+n+2}(e^{nr},e^{nr})
\label{29a}
\end{eqnarray}
As follows from Eqs. (\ref{28}) and (\ref{28a}), 
the possible values of $n$ are $n=0,1,2,...$ while,
as follows from Eqs. (\ref{29}) and (\ref{29a}), 
$r$ can take only the values of $0,1,....s$
(and therefore $s$ indeed has the meaning of the 
particle spin). 
Since $e^{nr}$ is the maximal $su(2)\times su(2)$ vector with the
eigenvalues of the operators ${\bf J}'$ and ${\bf J}"$ equal to
$n+r$ and $n+s-r$, respectively, then
as a basis of the representation space one can take the vectors
$e_{ij}^{nr}=(J_-')^i (J_-")^je^{nr}$
where, for the given $n$ and $s$, the quantity $i$ can
take the values of $0,1,...n+r$ and $j$ - the values 
of $0,1,...n+s-r$. 

In the subsequent section we consider in detail the
spinless case and show that the basis discussed 
in this section
is an implementation of the generators (\ref{12}) but
not (\ref{9}) and (\ref{10}). If $s=0$ then there
exist only the maximal $su(2)\times su(2)$ vectors $x^{kl}$
with $k=l$ and therefore the basis of the  representation  space  is
formed by the vectors $e_{\alpha\beta}^n\equiv e_{\alpha\beta}^{n0}$
where $n=0,1,2,...$; $\alpha,\beta=0,1,...n$. The explicit 
expressions for the  action  of  operators $R_{ij}$ in this
basis can be calculated by using Eq. (\ref{15}), and the
result is
\begin{eqnarray}
&& R_{11}e_{\alpha\beta}^n=\frac{(n+1-\alpha)(n+1-\beta)}
{(n+1)^2}e_{\alpha\beta}^{n+1}+ \nonumber\\
&&\frac{\alpha\beta n}{4(n+1)}
[w+(2n+1)^2]e_{\alpha -1,\beta-1}^{n-1},\nonumber\\
&&R_{12}e_{\alpha\beta}^n=\frac{n+1-\alpha}{(n+1)^2}
e_{\alpha,\beta+1}^{n+1}-\frac{\alpha n}{4(n+1)}[w+(2n+1)^2]
e_{\alpha -1,\beta}^{n-1},\nonumber\\
&& R_{21}e_{\alpha\beta}^n=\frac{n+1-\beta}{(n+1)^2}
e_{\alpha+1,\beta}^{n+1}-\frac{\beta n}{4(n+1)}[w+(2n+1)^2]
e_{\alpha,\beta-1}^{n-1},\nonumber\\
&&R_{22}e_{\alpha\beta}^n=\frac{1}{(n+1)^2}
e_{\alpha+1,\beta+1}^{n+1}+\nonumber\\
&&\frac{n}{4(n+1)}[w+(2n+1)^2]e_{\alpha \beta}^{n-1}
\label{34}
\end{eqnarray}
As follows from Eqs. (\ref{21}) and (\ref{28a})
\begin{equation}
(e_{\alpha\beta}^n,e_{\alpha\beta}^n)=
\frac{(n!)^2\alpha ! \beta !}{4^n(n+1)(n-\alpha)!(n-\beta)!}
\prod_{j=1}^{n}  [w+(2j+1)^2]
\label{35}
\end{equation}

\section{Implementation of UIRs in the space of functions
on the SU(2) group}
\label{S9}

As already noted, the three dimensional unit sphere $S^3$
represents the group space of the SU(2) group. The
elements $u\in SU(2)$ are often parametrized by the
Euler angles as follows (see e.g. Ref. \cite{Vilenkin})
\begin{eqnarray}
&&u_{11}=cos(\theta/2)exp(i(\chi+\psi)/2)\quad
  u_{12}=isin(\theta/2)exp(i(\chi-\psi)/2)\quad\nonumber\\
&&u_{21}=u_{12}^*\quad u_{22}=u_{11}^*
\label{36}
\end{eqnarray}
where $^*$ is used to denote the complex conjugation.
On the other hand, since the matrix $u$ can also be represented as 
$u=u_4+i{\bf \sigma}{\bf u}$, where $({\bf u},u_4)$ are the
components of the four-dimensional vector belonging to $S^3$,
one can express these components in terms of the Euler
angles. Note that in mathematical literature usually the 
notation $\varphi$ instead of $\chi$ is used, but we reserve 
$\varphi$ to denote the conventional polar angle, which,
as follows from Eq. (\ref{36}), is equal to $(\psi-\chi)/2$.
Also in physical literature the notation $\theta$ has
another meaning than in Eq. (\ref{36}) since the
"true" azimuthal angle $\theta$ is defined in such a way 
that $cos\theta = u_z/|{\bf u}|$.
For simplicity we do not rename the $\theta$ in 
Eq. (\ref{36}).

As a result of direct calculations, the 
form of representation generators (\ref{12}) for the
SU(2)$\times$SU(2) subgroup of the dS group are as follows
\begin{eqnarray}
&&J_+'=iexp(-i\chi)(\frac{\partial}{\partial \theta}+
\frac{i}{sin\theta}\frac{\partial}{\partial \psi}-
ictg\theta\frac{\partial}{\partial \chi})\nonumber\\
&&J_-'=-iexp(i\chi)(-\frac{\partial}{\partial \theta}+
\frac{i}{sin\theta}\frac{\partial}{\partial \psi}-
ictg\theta\frac{\partial}{\partial \chi})\nonumber\\
&&J_+"=-iexp(i\psi)(\frac{\partial}{\partial \theta}-
\frac{i}{sin\theta}\frac{\partial}{\partial \chi}+
ictg\theta\frac{\partial}{\partial \psi})\nonumber\\
&&J_-"=iexp(-i\psi)(-\frac{\partial}{\partial \theta}-
\frac{i}{sin\theta}\frac{\partial}{\partial \chi}+
ictg\theta\frac{\partial}{\partial \psi})\nonumber\\
&&J_z'=2i\frac{\partial}{\partial \chi}\quad
J_z"=-2i\frac{\partial}{\partial \psi}
\label{37}
\end{eqnarray}

Let $e_{\alpha\beta}^n(\chi,\theta,\psi)$ be a
function corresponding to the basis element 
$e_{\alpha\beta}^n$ in the preceding section for
a spinless UIR. Since $e_{\alpha\beta}^n$ is the
eigenvector of $J_3'$ with the eigenvalue $n-2\alpha$
and the eigenvector $J_3'$ with the eigenvalue $n-2\beta$,
then, as follows from Eq. (\ref{37}), the dependence of   
$e_{\alpha\beta}^n(\chi,\theta,\psi)$ on $\chi$
and $\psi$ is in the form
\begin{equation}
e_{\alpha\beta}^n(\chi,\theta,\psi)=
exp[-i(\rho\chi +\nu\psi)]
e_{\alpha\beta}^n(\theta)
\label{38}
\end{equation}
where $\rho = (n-2\alpha)/2$, $\nu = -(n-2\beta)/2$.

As noted in the preceding section, 
$e_{\alpha\beta}^n$ is also the eigenvector of
the operators ${\bf J}^{'2}$ and ${\bf J}^{"2}$
with the eigenvalue $n(n+2)$. We denote $l=n/2$
and $z=cos\theta$. Then, as
follows from Eqs. (\ref{37}) and (\ref{38})  
\begin{eqnarray}
&&(1-z^2)\frac{d^2e_{\alpha\beta}^n(z)}{dz^2}-2z
\frac{de_{\alpha\beta}^n(z)}{dz}-
\frac{\rho^2+\nu^2-2\rho\nu z}{1-z^2}e_{\alpha\beta}^n(z)
=\nonumber\\
&&-l(l+1)e_{\alpha\beta}^n(z)
\label{39}
\end{eqnarray} 
This is the equation for 
functions $P_{\rho\nu}^l(z)$ describing matrix elements 
of the regular representation of the SU(2) group
(see e.g. Ref. \cite{Vilenkin}). One of the convenient
forms of $P_{\rho\nu}^l(z)$ is as follows 
\cite{Vilenkin}
\begin{eqnarray}
&&P_{\rho\nu}^l(\theta)=i^{-\rho-\nu}
[\frac{(l-\rho)!(l-\nu)!}{(l+\rho)!(l+\nu)!}]^{1/2}
(ctg\gamma)^{\rho+\nu}\nonumber\\
&&\sum_{j=max(\rho,\nu)}^{l}
\frac{(l+j)!i^{2j}}{(l-j)!(j-\rho)!(j-\nu)!} 
(sin\gamma)^{2j}
\label{P}
\end{eqnarray}
where $\gamma = \theta/2$.
In some cases it is convenient to use the relation
between the functions $P_{\rho\nu}^l(z)$ and the
Jacobi polynomials \cite{Vilenkin}
\begin{eqnarray}
&&P_k^{(i,j)}=2^{\rho}i^{\nu-\rho}
[\frac{(l-\nu)!(l+\nu)!}{(l-\rho)!(l+\rho)!}]^{1/2}
(1-z)^{(\nu - \rho)/2}\nonumber\\
&&(1+z)^{-(\nu + \rho)/2}P_{\rho\nu}^l(z)
\label{40}
\end{eqnarray}
where
\begin{equation}
l=k+\frac{i+j}{2}\quad \rho = \frac{i+j}{2}\quad 
\nu = \frac{j-i}{2}
\label{41}
\end{equation}

We conclude that 
\begin{equation}
e_{\alpha\beta}^n(\chi,\theta,\psi)=c_{\alpha\beta}^n
exp[-i(\rho\chi +\nu\psi)]P_{\alpha\beta}^l(z)
\label{42}
\end{equation}
and our next task is to find the coefficients  
$c_{\alpha\beta}^n$. As follows from Eq. (\ref{37}),
\begin{eqnarray}
&&J_-'=-ie^{i\chi}[(1-z^2)^{1/2}\frac{\partial}{\partial z}
+\frac{i}{(1-z^2)^{1/2}}\frac{\partial}{\partial \psi}-
\frac{iz}{(1-z^2)^{1/2}}\frac{\partial}{\partial \chi}]\nonumber\\
&&J_-"=ie^{-i\psi}[(1-z^2)^{1/2}\frac{\partial}{\partial z}
-\frac{i}{(1-z^2)^{1/2}}\frac{\partial}{\partial \chi}+\nonumber\\
&&\frac{iz}{(1-z^2)^{1/2}}\frac{\partial}{\partial \psi}]
\label{43}
\end{eqnarray}
By using the relations \cite{Vilenkin}
\begin{eqnarray}
&&(1-z^2)^{1/2}\frac{\partial P_{\rho\nu}^l(z)}{\partial z}-
\frac{\rho z - \nu}{(1-z^2)^{1/2}}P_{\rho\nu}^l(z)=\nonumber\\
&&-i[(l+\rho)(l-\rho+1)]^{1/2}P_{\rho-1,\nu}^l(z)\nonumber\\
&&(1-z^2)^{1/2}\frac{\partial P_{\rho\nu}^l(z)}{\partial z}+
\frac{\nu z - \rho}{(1-z^2)^{1/2}}P_{\rho\nu}^l(z)=\nonumber\\
&&-i[(l+\nu)(l-\nu+1)]^{1/2}P_{\rho,\nu-1}^l(z)
\label{44}
\end{eqnarray}
and the properties 
$J_-'e_{\alpha\beta}^n=e_{\alpha+1,\beta}^n$ and
$J_-"e_{\alpha\beta}^n=e_{\alpha,\beta+1}^n$, we
now conclude that
\begin{equation}
e_{\alpha\beta}^n(\chi,\theta,\psi)=c_n (-1)^{\alpha}
[\frac{\alpha !\beta !}{(n-\alpha)!(n-\beta)!}]^{1/2}
exp[-i(\rho\chi +\nu\psi)]P_{\rho\nu}^l(z)
\label{45}
\end{equation}

Our final goal in this section is to find the
coefficients $c_n$ from the fact that the maximal
su(2)$\times$su(2) vectors $e^n$ are defined in
such a way that $R_{11}e^n=e^{n+1}$. Since 
$e^n==e_{00}^n$ then it follows from Eq. (\ref{45})
that $e^n$ depends on $\chi$ and $\psi$ only via
$exp(in\varphi)$ and on $z$ only via
$P_{l,-l}^l(z)$. We use the fact that \cite{Vilenkin}
\begin{equation}
P_{l,-l}^l(z)=i^{2l}(\frac{1-z}{2})^l
\label{46}
\end{equation}
Then, as follows from Eqs. (\ref{12}), (\ref{17}) 
and (\ref{36}), the action of $R_{11}$ on functions
depending on $\chi$ and $\psi$ only via 
$\varphi$ is given by
\begin{eqnarray}
R_{11}=e^{i\varphi}[cos\gamma\frac{\partial}{\partial \gamma}+
\frac{i}{sin\gamma}\frac{\partial}{\partial \varphi}+
\frac{i}{2}(\mu + 3i)sin\gamma]
\label{47}
\end{eqnarray}

Our conclusion is that
if $e^0(\chi,\theta,\psi)=1$ then the final
expression for $e_{\alpha\beta}^n(\chi,\theta,\psi)$
can be written in the form
\begin{eqnarray}
&&e_{\alpha\beta}^n(\chi,\theta,\psi)=\frac{n!}{2^n}
(-1)^{\alpha}[\frac{\alpha !\beta !}{(n-\alpha)!(n-\beta)!}]^{1/2}
\{\prod_{j=1}^{n}[\mu+i(2j+1)]\}\nonumber\\
&&exp[-i(\rho\chi +\nu\psi)]P_{\rho\nu}^l(cos\theta)
\label{48}
\end{eqnarray} 
where $\rho = (n-2\alpha)/2$, $\nu = -(n-2\beta)/2$, $l=n/2$.
The norm of $e_{\alpha\beta}^n$ defined in such a way is
compatible with Eq. (\ref{35}) since the volume element on
the SU(2) group in terms of the Euler angles is \cite{Vilenkin}
\begin{equation}
dV = \frac{1}{16\pi^2}sin\theta d\chi d\theta d\psi
\label{volume}
\end{equation}
$0\leq \chi <2\pi$, $0<\theta < \pi$, $-2\pi \leq \psi < 2\pi$
and 
\begin{equation}
\int |P_{\rho\nu}^l(cos\theta)|^2sin\theta d\theta = 
\frac{2}{2l+1}
\label{Pnorm}
\end{equation}

\section{Matrix elements in quasiclassical approximation}
\label{S10}

As follows from Eqs. (\ref{13}) and (\ref{49}), 
in the representation when the generators are given by 
Eq. (\ref{12}) the quasiclassical wave function can be
written as
\begin{equation}
f({\bf u})=a({\bf u}/u_4)u_4^{-3/2}
exp(imRlnu_4-im{\bf u}{\bf r}_0/u_4)
\label{51}
\end{equation}
since $\mu =2mR$ (see Sect. \ref{S5}). Note that if 
$R$ is very large then the exponent in this expression
is a rapidly oscillating function of $u_4$. Therefore,
in that case the dependence on $u_4$ is always
quasiclassical even if the dependence on the other 
variables is not. 

Instead of the Euler angles, one can also use the
following parametrization of the points on $S^3$:
\begin{eqnarray}
&&u_1=sin\gamma cos\varphi \quad 
u_2=sin\gamma sin\varphi\nonumber\\
&&u_3=cos\gamma cos\delta\quad
u_4=cos\gamma sin\delta
\label{52}
\end{eqnarray}
where $\varphi,\delta \in [0,2\pi]$ and $\gamma \in [0,\pi /2]$.
Then, as follows from Eq. (\ref{48}), the coefficient
$c_{\alpha\beta}^n$ defining the probability of the
state (\ref{51}) to have the quantum numbers 
$(n\alpha\beta)$ is given by
\begin{eqnarray}
&&c_{\alpha\beta}^n=\int exp[imRln(cos\gamma sin\delta) -
imv_0(x_0sin\gamma cos\varphi +y_0sin\gamma sin\varphi+\nonumber\\
&&z_0cos\gamma cos\delta)-i(n-\alpha -\beta)\varphi - 
i(\beta - \alpha)\delta)]\nonumber\\
&&P_{\rho\nu}^l(\gamma)
f_1(\varphi,\gamma,\delta)d\varphi d\delta d\gamma
\label{53}
\end{eqnarray}
where $f_1$ {\it is not} a rapidly oscilating function of its
variables. Here $(x_0,y_0,z_0)$ are the components of
${\bf r}_0$.

As follows from Eq. (\ref{17}), the $z$ component of the
angular momentum operator is given by $J_z=J_z'+J_z"$
and the operator $M_4^3$, which is equal to $2RP_z$
in the Poincare limit, is given by $M_4^3=J_z'-J_z"$.
Therefore the state $e_{\alpha\beta}$ is the eigenvector
of the operator $J_z$ with the eigenvalue $2n-2\alpha-2\beta$
and the eigenvector of the operator $M_4^3$ with the 
eigenvalue $2\beta-2\alpha$. We will now show that these
results are compatible with the quasiclassical approximation.
Indeed, the integral over $\varphi$ in Eq. (\ref{53})
has a stationary point if
\begin{equation}
m(x_0v_y-y_0v_x) = n-\alpha-\beta
\label{54}
\end{equation}
and the integral over $\delta$ in Eq. (\ref{53})
has a stationary point if
\begin{equation}
2mRv_z = 2(\beta-\alpha)
\label{55}
\end{equation}
The result (\ref{54}) is compatible with the quasiclassical
approximation since in our units $\hbar /2 =1$.

As follows from Eq. (\ref{55}), the maximum value
of $v_z$ is equal to $n/mR$. For this reason one might
think that the meaning of the quantum number $n$ when
$R$ is large is such that $|{\bf v}|=n/(mR)$. To prove
that this is the case one has to use asymptotic
expressions for $P_{\rho\nu}^l$ when the numbers
$(l\rho\nu)$ are large. As seen from Eq. (\ref{P}),
in the general case the functions $P_{\rho\nu}^l$ 
contain many oscillating terms and therefore the 
problem of finding their asymptotic expressions is
not easy. As follows from Eqs. (\ref{40})
and (\ref{41}), one can rewrite Eq. (\ref{53}) in the form
\begin{eqnarray}
&&c_{\alpha\beta}^n=\int exp[imRln(cos\gamma sin\delta) -
imv_0(x_0sin\gamma cos\varphi +\nonumber\\
&&y_0sin\gamma sin\varphi+
z_0cos\gamma cos\delta)-i(n-\alpha -\beta)\varphi - 
i(\beta - \alpha)\delta)]\nonumber\\
&&P_{N}^{(k_1,k_2)}(\gamma)sin^{k_1}(\gamma)cos^{k_2}(\gamma)
f_1(\varphi,\gamma,\delta)d\varphi d\delta d\gamma
\label{56}
\end{eqnarray}
where 
\begin{equation}
\label{57}
k_1=|n-\alpha-\beta|\quad k_2=|\beta-\alpha|\quad
N=(n-k_1 -k_2)/2
\end{equation}
One of the cases when it is possible to obtain
the asymptotic expression is when $N$ is large
while $k_1$ and $k_2$ are fixed (in particular this
means that $|v_z|<<|{\bf v}|$). In that case the 
result is (see e.g. Ref. \cite{BE})
\begin{eqnarray}
&&P_{N}^{(k_1,k_2)}(\gamma)=cos[(n+1)\gamma-
(2k_1+1)\pi/4]\nonumber\\
&&[(\pi N)^{1/2}(sin\gamma)^{k_1+1/2}
(cos\gamma)^{k_2+1/2}]^{-1}
\label{58}
\end{eqnarray}
As follows from this expression, the integral over
$\gamma$ in Eq. (\ref{56}) has a stationary point if
$n\approx mR|{\bf v}|$. We will see in subsequent 
sections that this result is valid in the general 
case, i.e. when $|v_z|$ is not necessarily small. 

\section{Free two-body mass operator}
\label{S11}

In contrast with Sect. \ref{S6}, we now assume that the
two-body system is free if its generators are sums of the
single-particle generators in the form (\ref{12})
(see the discussion in Sect. \ref{S6}). This implies
that $M_{ab}=M_{ab}^{(1)}+M_{ab}^{(2)}$
where $M_{ab}^{(1)}$ are the generators for the first 
particle and
$M_{ab}^{(2)}$ - for the second one. Each generator acts 
over the variables of its "own" particle, as described  
in Sect. \ref{S8}, and over the variables of another 
particle it acts as the identity operator. In other
words, the representation describing the two-body
system is the tensor product of single-particle UIRs.

 Denote by $\mu_1$  and $\mu_2$  $(\mu_1,\mu_2 >0)$ the dS
masses of the corresponding particles and assume that they are
spinless. Then, as follows from Eq. (\ref{23}), 
$I_2^{(1)}=2(w_1+9)$,
$I_2^{(2)}=2(w_2+9)$, where $w_1=\mu_1^2$,  $w_2=\mu_2^2$.
The tensor product of UIRs can be decomposed into the direct 
integral of UIRs and there exists a well elaborated general
theory \cite{Naimark}. In terminology of the theory of
induced UIRs, UIRs discussed in Sect. \ref{S4} belong to
the principal series of UIRs. In general, the decomposition
of the tensor product of UIRs belonging to the principal
series may contain not only UIRs of the principal series
(i.e. it may also contain UIRs not having "rest states"
defined by Eq. (\ref{26})). We will consider only a part
of the tensor product containing the "rest states"
and show that even for this part the spectrum of the
mass operator is not bounded below by the value of
$(\mu_1+\mu_2)^2$.

\begin{sloppypar}
It is clear that only UIRs with $s=0,2,4...$ 
(in our system of units) can enter the  tensor
product of two spinless representations. Therefore in 
order to find which values of $w$ are possible for the 
given $s$ one can act as
follows.  Construct $H_s$ ---  a space  of elements $x$,
satisfying the condition (compare with Eq. (\ref{26}))
\begin{equation}
 {\bf J}'x=J_+"x=0,\quad   {\bf J}^{"2} x=s(s+2)x
\label{59}
\end{equation}
Since $I_2$ commutes with all the representation operators
then $H_s$ is invariant under the action of $I_2$. 
Let $I_2^s$ be the reduction of $I_2$  onto $H_s$ . Define 
the operator W such that its reduction onto $H_s$ --- $W^s$
is defined from the relation $I_2^s =2[W^s -s(s+2)+9]$. 
Then the spectrum  of the operator $W^s$  defines the 
possible values of $w$ for the given $s$.
\end{sloppypar}

Note that although $W$ is the dS analog of the mass
operator squared in Poincare invariant theory, it is
not a square of any operator. Therefore one cannot exclude
a possibility that $W$ has even a negative part of the
spectrum. However, the part of $W$ corresponding to 
principles series UIRs has only the positive spectrum.

To construct a basis in the space $H_s$, we have first
to ascertain which linear  combinations  of  the  elements
$e_{\alpha_1\beta_1}^{(1)n_1}e_{\alpha_2\beta_2}^{(2)n_2}$
belong to $H_s$. Since $e_{\alpha_1\beta_1}^{(1)n_1}$ is 
the spinor with the spin $n_1$ with respect to the 
algebra ${\bf J}^{'(1)}$ as well as to ${\bf J}^{"(1)}$,
and  analogously for $e_{\alpha_2\beta_2}^{(2)n_2}$,  
then zero  eigenvalues of  the  operator ${\bf J}'$  can  be
obtained only if $n_1=n_2$, and the value s for the spin 
relative to the ${\bf J}"$ algebra can be obtained only if 
$n_1,n_2\geq s/2$. The
explicit finding of the required linear combinations can 
be performed by using Clebsch-Gordan coefficients for the 
$su(2)\times su(2)$ algebra but it is also possible
to verify directly that the vectors
\begin{eqnarray}
\Phi_n&=& \frac{4^{n+j}(1+2j)!}{[(n+j)!]^2j!}
\sum_{\alpha=0}^{n+j}\sum_{\beta=0}^n
(-1)^{\alpha+\beta}\times\nonumber\\
&&\frac{(j+\beta)!(n+j-\beta)!}{\beta!(n-\beta)!}
e_{\alpha\beta}^{(1)n+j}
e_{n+j-\alpha,n-\beta}^{(2)n+j}
\label{60}
\end{eqnarray}
where $j=s/2$ and $n=0,1,2...$, belong to $H_s$.
The value of $j$ is obviously equal to the spin
of the two-body system in conventional units. The
fact that the vectors $e^{(1)}$ and $e^{(2)}$
enter Eq. (\ref{60}) with the same value of the
quantum number $n$ supports the interpretation
of $n/R$ as the magnitude of the momentum 
(see the preceding section) since, by analogy
with Poincare invariant theory, one would
expect that the magnitudes of particle momenta in 
their common c.m. frame are equal to each other.

It is obvious that the  vectors
$\Phi_n$   with different $n'$s are orthogonal to each other.
The result of the calculation of the norm of the vector $\Phi_n$
(see Ref. \cite{lev3} for details) is
\begin{eqnarray}
(\Phi_n,\Phi_n)&=&\{\prod_{l=1}^{n+j} [w_1+(2l+1)^2]
[w_2+(2l+1)^2]\}\times\nonumber\\
&&\frac{(n+2j+1)!(1+2j)!}{(n+j+1)n!}
\label{61}
\end{eqnarray}
(we have fixed a misprint in Eq. (32) of Ref. \cite{lev3}).

     Our next goal is to find how the operator $W^s$  acts 
in the basis $\{\Phi_n\}$. As follows from Eq. (\ref{22}),
\begin{eqnarray}
&&I_2^s \Phi_n=8[(R_{22}^{(1)}+R_{22}^{(2)})(R_{11}^{(1)}+
R_{11}^{(2)})-(R_{21}^{(1)}+R_{21}^{(2)}) \times\nonumber\\
&&(R_{12}^{(1)}+R_{12}^{(2)})]\Phi^n -4s(s+2)\Phi_n,
\label{62}
\end{eqnarray}
\begin{eqnarray}
&&I_2^{(1)}e_{\alpha\beta}^{(1)n}=
8(R_{22}^{(1)}R_{11}^{(1)}-R_{21}^{(1)}R_{12}^{(1)})
e_{\alpha\beta}^{(1)n}-8[n(n+2)+ \nonumber\\
&&(n-2\alpha)]e_{\alpha\beta}^{(1)n}=2(w_1 +9)e_{\alpha\beta}^{(1)n}
\label{63}
\end{eqnarray}
and an analogous formula takes place for
$I_2^{(2)}e_{\alpha\beta}^{(2)n}$.

Taking into account Eqs. (\ref{34}), (\ref{60}), (\ref{62}),
(\ref{63}) and the definition of the operator $W^s$, a
direct calculation shows that
\begin{equation}
  W^s\Phi_n=\sum_{l=0}^{\infty}\Phi_l  W_{ln}^s
\label{64}
\end{equation}
where the matrix $||W_{ln}^s||$ has only the following components
different from zero:
\begin{eqnarray}
&& W_{n+1,n}^s=\frac{n+1}{n+1+j},\quad
W_{nn}^s=w_1+w_2 +8(n+1)^2+ \nonumber\\
&&2s(4n+3)+s^2 +1,\quad W_{n,n+1}^s=\frac{n+2j+2}
{n+j+2}\times\nonumber\\
&&[w_1+(2n+2j+3)^2][w_2+(2n+2j+3)^2]
\label{65}
\end{eqnarray}
Such a matrix is called three-diagonal and in fact, 
only the terms with $l=n-1,n,n+1$
contribute to the sum (\ref{64}). Note that
the operator $W^s$   is  certainly  Hermitian,  but  
since  the  basis elements are not normalized to one, 
the Hermiticity  condition
has not the usual form $W_{nl}^s=W_{ln}^{s*}$, but
$||\Phi_n||^2W_{nl}^s=||\Phi_l||^2W_{ln}^{s*}$.

The matrix of the operator $W^s-\lambda$ has the matrix elements
$||W_{nl}^s -\lambda\delta_{nl}||$. We use $\Delta^n(\lambda)$
to denote the determinant of the matrix obtained from this
one by taking into account only the rows and columns with the
numbers $0,1,...n$. It  is well known (and can be verified 
directly)  that  for  the  three-diagonal matrix
the following relation is valid:
\begin{equation}
\Delta^{n+1}(\lambda)=(W_{n+1,n+1}^s-\lambda)\Delta^n(\lambda)-
W_{n+1,n}^sW_{n,n+1}^s \Delta^{n-1}(\lambda)
\label{66}
\end{equation}
where it is formally assumed  that $\Delta^{-1}(\lambda)=1$. 

Since we consider only representations of the  principle series,
we are interesting only in the region of positive
$\lambda$'s. Therefore we can represent $\lambda$ as
$\lambda=(\mu_1+\mu_2+\sigma)^2$  where
$\sigma$ has the meaning of the dS kinetic energy. However as 
we will see below, not  only  $\sigma\geq 0$,
but also $\sigma <0$ is possible.

Consider the vector
\begin{equation}
\chi_{\lambda}=\sum_{n=0}^{\infty} (-1)^n
\Delta^n(\lambda)[\prod_{l=1}^nW_{l,l+1}^s]^{-1}\Phi_n
\label{67}
\end{equation}

As follows from Eqs. (\ref{65}) and (\ref{66}), it is formally
an  eigenvector  of  the
operator $W^s$  with the eigenvalue $\lambda$. If the sum in
Eq. (\ref{67})  converges,
i.e. $(\chi_{\lambda},\chi_{\lambda})$ is finite, 
then $\lambda$ is
the true eigenvalue i.e. it belongs
to the discrete spectrum of the operator $W^s$. It is also possible
that $\chi_{\lambda}$  is the generalized eigen vector, i.e.
$(\chi_{\lambda},\chi_{\lambda'})$ is proportional to
$\delta(\lambda-\lambda')$. Then $\lambda$ belongs to the
continuous  spectrum  of the operator $W^s$. A detailed 
calculation has been carried out in Ref. \cite{lev3} and the 
result is 
\begin{eqnarray}
&& (\chi_{\lambda},\chi_{\lambda'})=\frac{\pi^{1/2}\Gamma((3+s)/2)}
{2^s(1+s/2)(1+s/2)!}\{
\prod_{j=1}^{s/2}[w_1+(2j+1)^2]\nonumber\\
&&[w_2+(2j+1)^2]\}\delta(\lambda-\lambda')
|\Gamma((3+s-\imath \mu_1)/2)\nonumber\\
&&\Gamma((3+s-\imath \mu_2)/2)
\Gamma((\imath/2)(\mu_1+\mu_2+\sigma))|^2\nonumber\\
&&|\Gamma([3+s-\imath(2\mu_1+2\mu_2+\sigma)]/4)
\Gamma((3+s-\imath \sigma)/4)\nonumber\\
&&\Gamma([3+s-\imath(2\mu_1+\sigma)]/4)\nonumber\\
&&\Gamma([3+s-\imath(2\mu_2+\sigma)]/4)|^{-2}
\label{68}
\end{eqnarray}
where $\Gamma$ is the gamma function.
Therefore the discrete spectrum is absent and
continuous one fills all the interval $\lambda\in (0,\infty)$.

In Ref. \cite{lev1}, where the explicit expression 
for the two-body mass operator in the form  of
a differential operator in some space  of  functions 
has been has been found, a similar result has been obtained. 
The result (\ref{68}), however (obtained in Ref. \cite{lev3}) 
is in fact algebraic since the operator $W^s$ has been 
considered in the form of an infinite matrix. In Sect.
\ref{S6} we have shown that the dS mass operator has
the infinite spectrum in the range $(-\infty,\infty)$.
However, this result has been obtained in the nonrelativistic
approximation and in first order in $1/R$. On the contrary,
no approximation has been assumed in deriving Eq. (\ref{68}).

\section{Internal Hilbert space for the two-body system}
\label{S12}

When a two-body system is considered in Galilei and 
Poincare invariant theories, it is often convenient to
describe its wave function not only in terms of
individual particle variables but also in terms of
external and internal variables. External variables
describe the motion of the system as a whole while
internal variables describe the relative motion. In
Galilei invariant theories the two-body mass operator
acts only in the Hilbert space of wave functions
describing the relative motion. In Poincare invariant
theories the two-body mass operator is unitarily
equivalent to one acting only in the internal 
Hilbert space (see e.g. Ref. \cite{Sokolov,rqm}). The
problem arises what is the internal Hilbert space
in dS invariant theories. 

Let us consider the vector $\Phi_n$
(see Eq. (\ref{60})) in the case when the vectors
$e_{\alpha\beta}^n$ are treated as functions on the
SU(2) group space (see Eq. (\ref{48})). We denote
\begin{equation} 
T_{\rho\nu}^l(u)=exp[-i(\rho \chi + \nu \psi)] 
P_{\rho\nu}^l(\theta)
\label{69}
\end{equation}
where $u\in SU(2)$ is defined by the Euler angles
$(\chi \theta \psi)$. Eq. (\ref{69}) defines the
matrix of the representation operator $T^l$ in the
case when the regular representation of the SU(2)
group is reduced onto the subspace of functions
which are the eigenfunctions of the operators
${\bf J}^{'2}$ and ${\bf J}^{"2}$ with the
eigenvalue $n(n+2)$, $n=2l$ (see e.g. Ref. 
\cite{Vilenkin}). The functions $e^{(1)}$ and
$e^{(2)}$ depend on the corresponding single-particles
variables, which we denote as $u_1$ and $u_2$,
respectively. Then, as follows from Eqs. (\ref{48}),
(\ref{60}) and (\ref{69})
\begin{eqnarray}
&&\Phi_n(u_1,u_2)=\frac{(-1)^{n+j}}{j!}
\{\prod_{k=1}^{n+j}[\mu_1+i(2k+1)][\mu_2+i(2k+1)]\}\nonumber\\
&&\sum_{\alpha=0}^{n+j}\sum_{\beta =0}^n 
(-1)^{\alpha + \beta}[\frac{(j+\beta)!(n+j-\beta)!}
{\beta ! (n-\beta)!}]^{1/2}\nonumber\\
&&T_{\rho_1\nu_1}^{l+j/2}(u_1)T_{\rho_2\nu_2}^{l+j/2}(u_2)
\label{70}
\end{eqnarray}
where
\begin{eqnarray}
&&\rho_1=\frac{1}{2}(n+j-2\alpha)\quad 
\nu_1=- \frac{1}{2}(n+j-2\beta)\quad 
\rho_2=\frac{1}{2}(2\alpha - n-j)\nonumber\\
&&\nu_2=- \frac{1}{2}(2\beta+j-n)
\label{71}
\end{eqnarray}

Consider now the following question. 
If the $T_{\rho\nu}^l(u)$ are the matrix elements of
the representation operator $T^l(u)$ then what are
the matrix elements of the operator $T^l(u^{-1})$?
We use $g(\varphi)$ and $h(\theta)$ to denote the
following SU(2) elements
\begin{equation}
g(\chi)=
\left\|\begin{array}{cc}
exp(i\chi/2)&0\\
0&exp(-i\chi/2)
\end{array}\right\|\quad 
h(\theta)=\left\|\begin{array}{cc}
cos(\theta/2)&isin(\theta/2)\\
isin(\theta/2)&cos(\theta/2)
\end{array}\right\|
\label{72}
\end{equation}
If $u\in SU(2)$ is characterized by the Euler angles
$(\chi\theta\psi)$ then 
$u=g(\chi)g(\theta)g(\psi)$ and therefore
$u^{-1}=g(-\psi)g(-\theta)g(-\chi)$. Then, as follows
from Eq. (\ref{69})
\begin{equation} 
T_{\rho\nu}^l(u^{-1})=exp[i(\rho \psi + \nu \chi)] 
P_{\rho\nu}^l(-\theta)
\label{73}
\end{equation}
Note that $P_{\rho\nu}^l(-\theta)$ is not defined yet
since the argument should be in the range $(0,\pi)$.
We can use the fact \cite{Vilenkin} that 
$h(-\theta)=g(\pi)h(\theta)g(-\pi)$.
Then, as follows from Eqs. (\ref{69}) and (\ref{73})
\begin{equation} 
T_{\rho\nu}^l(u^{-1})=(-1)^{\nu -\rho}T_{-\nu,-\rho}^l(u)
\label{74}
\end{equation}

As follows from Eq. (\ref{74}) and the property of
representation operators $T^l(u_1u_2)=T^l(u_1)T^l(u_2)$,
Eq. (\ref{70}) can now be written in the form
\begin{eqnarray}
&&\Phi_n(u_1,u_2)=\frac{(-1)^j}{j!}
\{\prod_{k=1}^{n+j}[\mu_1+i(2k+1)][\mu_2+i(2k+1)]\}\nonumber\\
&&\sum_{\nu = -l}^l[\frac{(l+j+\nu)!(l+j-\nu)!}
{(l+\nu) ! (l-\nu)!}]^{1/2}
T_{\nu +j/2, \nu -j/2}^{l+j/2}(u_2^{-1}u_1)
\label{75}
\end{eqnarray}

We conclude that the function $\Phi_n$ describing an
internal state, actually depends not on single-particle
variables $u_1$ and $u_2$ independently but only on
their combination $u_2^{-1}u_1$. Therefore the latter
can be treated as a variable describing the internal
motion of the two-body system. This is a generalization
of the well known fact that in nonrelativistic quantum
mechanics the internal momentum ${\bf q}$ is defined
as in Eq. (\ref{AG2}). Indeed, when the velocities are
small, $u_1$ is defined by the three-dimensional vector
${\bf u}_1\approx {\bf v}_1$ such that 
$|{\bf u}_1|\ll 1$ and $u_1\approx 1 +i{\bf \sigma}{\bf u}_1$.
Analogously $u_2\approx 1 +i{\bf \sigma}{\bf u}_2$.
Therefore $u_2^{-1}u_1$ in this approximation is defined
by ${\bf u}={\bf u}_1-{\bf u}_2$. On the other hand, 
${\bf p}_k=m_k{\bf u}_k$, $(k=1,2)$ and therefore the
relative momentum ${\bf q}$ in the nonrelativistic
approximation is equal to $m_{12}{\bf u}$
where $m_{12}$ is the reduced mass. 

Let us stress that
our conclusion that $u_2^{-1}u_1$ is the internal
variable in dS invariant theories is obtained without
involving any approximation. For this reason it is
possible to derive the relation between ${\bf u}$ and
${\bf q}$ in the general case. Recall that 
${\bf u}_j={\bf v}_j/(v_0)_j$ ($j=1,2$) (see Sect. 
\ref{S4}). Therefore, in the c.m. frame, where 
${\bf p}_1={\bf q}$ and ${\bf p}_2=-{\bf q}$,
the elements of the SU(2) group corresponding to $u_1$ and 
$u_2$ are given by
\begin{equation}
u_1=\frac{m_1+i{\bf \sigma}{\bf q}}{(m_1^2+q^2)^{1/2}}\quad 
u_2=\frac{m_2-i{\bf \sigma}{\bf q}}{(m_2^2+q^2)^{1/2}} 
\label{qu1}
\end{equation}   
where $q=|{\bf q}|$. We now use $u$ to denote the internal 
variable $u_2^{-1}u_1$. Since the element of the SU(2) group
corresponding to $u$ is $u_4+i{\bf \sigma}{\bf u}$, 
it follows from Eq. (\ref{qu1}) that
\begin{equation}
{\bf u}=\frac{(m_1+m_2){\bf q}}{[(m_1^2+q^2)(m_2^2+q^2)]^{1/2}}\quad
u_4=\frac{m_1m_2-q^2}{[(m_1^2+q^2)(m_2^2+q^2)]^{1/2}}
\label{qu2}
\end{equation} 

The vector $\Phi_n$ in Eq. (\ref{75})
depends only on $u\in SU(2)$, and we can
parametrize $u$ by the Euler angles, for which we
will again use the notations $(\chi\theta\psi)$.
Then the internal polar angle is $\varphi = (\psi-\chi)/2$
(see Sect. \ref{S9}) and, as follows from Eq. (\ref{69}),
the dependendence of $\Phi_n$ on $\varphi$ is in
the form $exp(ij\varphi)$. This is in agreement with
the fact that $\Phi_n$ is the eigenvector of the $z$
component of the two-body spin operator ${\bf S}$ with 
the eigenvalue $j$ (in conventional units). 

Since $\Phi_n$ is also the eigenvector of ${\bf S}^2$ with
the eigenvalue $j(j+1)$, one might expect that $\Phi_n$
depends on $\varphi$ and the "true" azimuthal angle
${\tilde \theta}$ (see Sect. \ref{S9}) in the form
$Y_{jj}({\tilde \theta},\varphi)=
exp(ij\varphi)P_j^j({\tilde \theta})$, where $Y_{jj}$
is the spherical function and $P_j^j$ is the associated
Legendre function. However, since the main goal of the
present paper is to demonstrate the idea that 
gravitational effects can be obtained in the free
theory, we will concentrate our attention on the
simplest case when the internal two-body spin is
equal to zero, i.e. $j=0$. Then Eq. (\ref{75}) has a much 
simpler form: 
\begin{eqnarray}
\Phi_n(u)=\{\prod_{k=1}^n[\mu_1+i(2k+1)]
[\mu_2+i(2k+1)]\}\sum_{\nu = -l}^lT_{\nu\nu}^l(u)
\label{76}
\end{eqnarray}

The sum in Eq. (\ref{76}) is the trace of
the representation operator $T^l(u)$, and in the
literature it is called the character of the 
representation $u\rightarrow T^l(u)$. The character
can be calculated in a much simpler way than the
general result when $j\neq 0$ \cite{Vilenkin}. Define
a parameter $t$ such that
\begin{equation}
cost=cos\frac{\theta}{2}cos\frac{\chi+\psi}{2}
\label{77}
\end{equation}
Then
\begin{equation}
\sum_{\nu = -l}^lT_{\nu\nu}^l(u)=\frac{sin(n+1)t}{sint}
\label{78}
\end{equation}
As follows from Eq. (\ref{36}), $cost=u_4$ and therefore
$t$ has a clear meaning: since $u_4^2+{\bf u}^2=1$,
$t$ is the angle between the four-dimensional unit
vector $u$ and the 4 axis. In particular, the result
(\ref{78}) shows that the character depends only on
$|{\bf u}|$ and does not depend on 
$\varphi$ and ${\tilde \theta}$, as expected.

We conclude that in the case when the internal
angular momentum of the two-body system is zero,
the functions $\Phi_n$ have the following simple form: 
\begin{eqnarray}
\Phi_n(u)=\{\prod_{k=1}^n[\mu_1+i(2k+1)]
[\mu_2+i(2k+1)]\}\frac{sin(n+1)t}{sint}
\label{79}
\end{eqnarray}
It is easy to show that the characters are normalized
to one \cite{Vilenkin} and therefore the 
normalization of states is compatible with Eq. (\ref{61}).

\chapter{Mean value of the two-body mass operator}
\label{Ch4}

\section{Preliminary discussion}
\label{prelim}

Let $\Psi(u_1,u_2)$ be a wave function of the two-body
system and one wants to calculate the distribution of
possible values of the mass in the state $\Psi$.
In Poincare invariant theory an analogous problem
has been investigated in different approaches 
\cite{Sokolov,rqm} and the results are similar. Namely,
instead of individual particle variables 
$({\bf p}_1,{\bf p}_2)$ one can introduce the total
momentum ${\bf P}$, the relative momentum ${\bf q}$
and decompose the two-body Hilbert space into a
direct integral of Hilbert spaces $H({\bf P})$
corresponding to a given value of ${\bf P}$. The
mass operator can be decomposed into a direct
integral of operators $M({\bf P})$ acting in
corresponding spaces $H({\bf P})$. The spectra
of all the operators $M({\bf P})$ are the same,
which is the manifestation of the fact that ${\bf P}$
is only a kinematical variable while dynamics is
fully defined by the mass operator. Therefore the
dynamics is fully defined by the operator $M(0)$
in $H(0)$. In turn, since $M(0)$ commutes with
the two-body spin operator ${\bf S}$, its spectrum
can be investigated by decomposing $H(0)$ into
subspaces $H_{jm}(0)$ corresponding to the eigenvalue
of ${\bf S}^2$ equal to $j(j+1)$ and the eigenvalue
of $S_z$ equal to $m$.

In the preceding chapter we have investigated the
structure of the Hilbert space $H_s$, which is the
analog of $H_{jj}(0)$ in Poincare invariant theory.
By analogy with the construction of UIRs in Sect.
\ref{S8}, one can construct the full two-body 
Hilbert space as a direct sum of spaces identical
to $H_s$. Since the mass operator commutes with
all the representation operators, it also commutes
with all the projectors onto such spaces and its
spectra in each space are the same. Therefore to
investigate the spectrum of the two-body mass
operator it is sufficient to consider its spectrum
in $H_s$. As shown in Sect. \ref{S11}, the 
the operator $W$, which is the analog of the mass operator 
squared in Poincare invariant theory, has the spectrum 
containing the interval $(0,\infty)$ (and possibly
other points) i.e. the spectrum is not
bounded below by $(\mu_1+\mu_2)^2$ as one would
expect.
Therefore one has to investigate whether the
states having the values of the mass less than
$\mu_1+\mu_2$ are physical or not. For this purpose
it is sufficient to investigate not all possible
wave functions $\Psi(u_1,u_2)$ but only their 
projections onto $H_s$.

Instead of the states $\Phi_n$ in $H_s$, we now
introduce the states
\begin{equation}
\Psi_n=\Phi_n(\Phi_n,\Phi_n)^{-1/2}
\label{80}
\end{equation}
Then the elements $\{\Psi_n\}$ form the orthonormal
basis in $H_s$. As follows from Eqs. (\ref{61}) and 
(\ref{65}), the nonzero matrix elements of 
the operator $W^s$ in the basis $\{\Psi_n\}$ are 
given by
\begin{eqnarray}
&& W_{n+1,n}=W_{n,n+1}=\{[w_1+(2n+2j+3)^2]\nonumber\\
&&[w_2+(2n+2j+3)^2]\frac{(n+2j+2)(n+1)}{(n+1+j)
(n+2+j)}\}^{1/2} \nonumber\\
&&W_{nn}=w_1+w_2+8(n+1)^2+4j(4n+3)+4j^2+1
\label{81}
\end{eqnarray} 
where $j=s/2$.

Let
\begin{equation}
\Psi = \sum_{n=0}^{\infty} c_n\Psi_n\quad 
(\sum_{n=0}^{\infty} |c_n|^2=1)
\label{82}
\end{equation}
be a state in $H_s$. Then the mean value of the
operator $W$ in the state $\Psi$ is given by
\begin{eqnarray}
&&(\Psi,W\Psi)=\sum_{n=0}^{\infty} 
[W_{nn}|c_n|^2 +2W_{n,n+1}Re(c_{n+1}c_n^*)]
\label{83}
\end{eqnarray}
where $W_{nn}$ and $W_{n,n+1}$ are given by Eq.
(\ref{82}). Since we are interested in comparing 
the results
with Poincare invariant theory, the mass operator
squared in Poincare terms should be defined as
$M^2=W/4R^2$ (see Sects. \ref{S5} and \ref{S11}). 
We also take into
account that typical values of $n$ are very large
since, as noted in the preceding chapter, 
$n$ is of order $qR$ where ${\bf q}$ is 
the relative momentum and $q=|{\bf q}|$. 
In particular, $n\gg j$ since
for quasiclassical particles $j$ is of order
$qr$ where $r$ is the distance between the
particles. Then taking into account the
relation between the dS and Poincare masses (see Sect.
\ref{S5}), we get from Eq. (\ref{83})
\begin{eqnarray}
&&(\Psi,M^2\Psi)=\sum_{n=0}^{\infty} 
\{[m_1^2+m_2^2+2(n/R)^2]|c_n|^2 +\nonumber\\
&&2[m_1^2+(n/R)^2]^{1/2}[m_2^2+(n/R)^2]^{1/2}
Re(c_{n+1}c_n^*)\}
\label{84}
\end{eqnarray}
i.e. the dependence on $s$ formally dissapears.

\section{Standard relativistic mass operator}
\label{relativistic}

Before proceeding to calculations, let us look at
this expression more carefully. If the particles are
quasiclassical then one would expect that the 
dependence on $n$ of the
coefficients $c_n$ has a maximum around $qR$ and
the width of the maximum is much less than $qR$.
Suppose also that the difference between the 
consecutive coefficients is small, i.e. 
$c_n\approx c_{n+1}$. Then it follows from Eq. (\ref{84})
that the mean value of the operator $M^2$ is given
by the standard relativistic expression for the free
mass operator:
\begin{equation}
{\bar M}^2=M_0^2=[(m_1^2+q^2)^{1/2}+(m_2^2+q^2)^{1/2}]^2
\label{85}
\end{equation} 

Suppose now that $c_n$ and $c_{n+1}$ have approximately
the same magnitudes but the difference between their
phases - $\delta$ - cannot be neglected. 
If $\delta$ does not change significantly when $n$ is
inside the width of the maximum then instead of Eq. 
(\ref{85}) we have 
\begin{eqnarray}
{\bar M}^2=M_0^2-4(sin\frac{\delta}{2})^2[m_1^2+q^2]^{1/2}
[m_2^2+q^2]^{1/2}
\label{86}
\end{eqnarray} 
The contribution of the last term in this expression is 
always negative. In the nonrelativistic approximation  
it is proportional to the particle masses. In particular, 
if $\delta \ll 1$ then in the nonrelativistic approximation
\begin{eqnarray}
{\bar M}^2=M_0^2-m_1m_2\delta^2
\label{87}
\end{eqnarray} 
where $M_0$ should also be taken in the nonrelativistic
approximation: $M_0=m_1+m_2+q^2/(2m_{12})$.

If Newtonian gravity is taken into account then the
nonrelativistic mass of the classical two-body system is
\begin{equation}
M = M_0 -\frac{Gm_1m_2}{r_0}
\label{88}
\end{equation}
where $r_0$ is the distance between the particles.
As a result, in the nonrelativistic approximation
\begin{equation}
M^2 = M_0^2 -\frac{2G(m_1+m_2)}{r_0}m_1m_2
\label{89}
\end{equation}
This expression is compatible with Eq. (\ref{87}) if
\begin{equation}
\delta = \delta_0(r_0)=[\frac{2G(m_1+m_2)}{r_0}]^{1/2}=
(\frac{r_{1G}+r_{2G}}{r_0})^{1/2}
\label{90}
\end{equation}
where $r_{1G}=2Gm_1$ and $r_{2G}=2Gr_2$. In the
literature the quantity $r_G=2Gm$ is called the 
gravitational radius of the particle with the mass $m$.
Therefore the condition $\delta \ll 1$ has a clear
meaning: the sum of the gravitational radii of the 
particles in the two-body system should be much less
than the distance between them. This is precisely the 
condition when Newtonian gravity is a good first 
approximation to General Relativity. Another
important observation is as follows. If 
gravity is a consequence of Eqs. (\ref{85}) and
(\ref{86}) then the gravitational potential is always 
nonsingular since at small values of $r$ the condition 
$\delta \ll 1$ is not satisfied.  

As noted in Sect. \ref{S3}, any dS invariant 
quantum theory can be constructed only in terms of
dimensionless quantities. In particular, in terms of
dS masses $\delta$ can be written as 
${\tilde G}(\mu_1+\mu_2)/\varphi$ where the angular
variable $\varphi$ can be treated as $r/(2R)$ and
${\tilde G}=G/(4R^2)$. However, since we wish to
compare our results with the standard ones, in this
chapter we will use the standard gravitational
constant and masses.

We now proceed to actual calculations. Consider a system
of two spinless quasiclassical particles and assume
that their wave functions can be taken as in Eqs. (\ref{49})
and (\ref{50}). We also assume (see the discussion in
Sect. \ref{S7}) that if the particles do not interact with 
each other then the two-body wave function is the product of 
the single-particle wave functions. As noted in the 
preceding chapter, all the information about the mass 
operator can be obtained by considering its action only in 
the spaces $H_s$. For this reason we consider only the case 
when the total momentum of the two-body system is equal to 
zero. 

We first consider the nonrelativistic approximation. 
As noted above, in that case ${\bf u}\approx {\bf v}$,
$|{\bf u}|\ll 1$ and hence $u_4\approx 1-{\bf u}^2/2$.
Therefore, taking into account Eqs. (\ref{49}) and
(\ref{50}), one can write the two-body wave function as
\begin{eqnarray}
&&\Psi({\bf u}_1,{\bf u}_2)=const\,\, exp\{-\frac{iR}{2}
(m_1{\bf u}_1^2+m_2{\bf u}_2^2)-
i(m_1{\bf u}_1{\bf r}_1+\nonumber\\
&&m_2{\bf u}_2{\bf r}_2)-
\frac{1}{2}[b_1^2({\bf u}_1-{\bf g}_1)^2+
b_2^2({\bf u}_2-{\bf g}_2)^2]\}
\label{91}
\end{eqnarray}
where we use ${\bf g}_1$ and ${\bf g}_2$ to denote the
quantity ${\bf v}_0$ for particles 1 and 2, respectively.

As noted in the preceding chapter, the quantity 
${\bf u}={\bf u}_1-{\bf u}_2$ has the meaning of the
relative velocity in the nonrelativistic approximation.
The question arises what is the correct choice of the
external two-body variables. Recall that for calculating
the mass distribution for the two-body system with zero
total momentum, one has to calculate the coefficients
$c_n=(\Psi_n,\Psi)$. Since $\Psi_n$ depends only on internal
variables, the dependence on external variables in the
integral for $c_n$ will be integrated out. Therefore,
any variable ${\bf U}$, such that there exists
a one-to-one relation between the sets $({\bf U}{\bf u})$ and
$({\bf u}_1{\bf u}_2)$, can be taken as the external 
variable. By analogy with the nonrelativistic theory
we choose ${\bf U}=(m_1{\bf u}_1+m_2{\bf u}_2)/(m_1+m_2)$
as the external velocity variable. Then
\begin{equation}
{\bf u}_1={\bf U}+\frac{m_2}{m_1+m_2}\quad
{\bf u}_2={\bf U}-\frac{m_1}{m_1+m_2}
\label{92}
\end{equation} 
and Eq. (\ref{91}) can be rewritten as 
\begin{eqnarray}
&&\Psi({\bf U},{\bf u})=const\,\, exp\{-\frac{iR}{2}
((m_1+m_2){\bf U}^2+m_{12}{\bf u}^2)-\nonumber\\
&&i((m_1+m_2){\bf U}{\bf R}+
m_{12}{\bf u}{\bf r})-
\frac{1}{2}[b_1^2({\bf U}+\frac{m_2}{m_1+m_2}-\nonumber\\
&&{\bf g}_1)^2+
b_2^2({\bf U}-\frac{m_1}{m_1+m_2}-{\bf g}_2)^2]\}
\label{93}
\end{eqnarray} 
where
\begin{equation}
{\bf R}=\frac{m_1{\bf r}_1+m_2{\bf r}_2}{m_1+m_2}\quad
{\bf r}={\bf r}_1-{\bf r}_2
\label{94}
\end{equation}
Therefore Eq. (\ref{93}) can be represented as
\begin{eqnarray}
&&\Psi({\bf U},{\bf u})=const\,\, 
exp\{-\frac{{\bf U}^2}{2}[b_1^2+b_2^2+i(m_1+m_2)R]+\nonumber\\
&&{\bf U}[b_1^2({\bf g}_1-
\frac{m_2{\bf u}}{m_1+m_2})+
b_2^2({\bf g}_2+\frac{m_1{\bf u}}{m_1+m_2})-\nonumber\\
&&i(m_1+m_2){\bf R}]\}\Psi({\bf u})
\label{95}
\end{eqnarray}
where
\begin{eqnarray}
&&\Psi({\bf u})=exp\{-\frac{i}{2}Rm{\bf u}^2-
\frac{b_1^2}{2}(\frac{m{\bf u}}{m_1}-{\bf g}_1)^2-\nonumber\\
&&\frac{b_2^2}{2}(\frac{m{\bf u}}{m_2}+{\bf g}_2)^2
-im{\bf u}{\bf r}\}
\label{96}
\end{eqnarray}  
and we write $m=m_{12}$ for simplicity.

Since our goal is to calculate $c_n$, and the vectors
$\Psi_n$ do not depend on the external variables, we
can integrate $\Psi({\bf U},{\bf u})$ over ${\bf U}$.
Since we assume that $R$ is very large, it is clear
from Eq. (\ref{95}) that 
\begin{equation}
\int \Psi({\bf U},{\bf u})d^3{\bf U}= const\,\, \Psi({\bf u})
\label{97}
\end{equation} 
where $const$ does not depend on the parameters characterizing
the quasiclassical wave functions, i.e. does not depend on
whether the functions $a({\bf v})$ in Eq. (\ref{49}) are taken
as in Eq. (\ref{50}) or in other forms. This result shows
that $\Psi({\bf u})$ is proportional to $\Psi(0,{\bf u})$. 
As a consequence, 
\begin{equation}
c_n= const\int \Psi({\bf u})\Psi_n({\bf u})^*d^3{\bf u}
\label{98}
\end{equation}
where $\Psi_n$ belongs to a certain $H_s$.

We will describe all the necessary steps in the case when the
internal two-particle spin is equal to zero. One might 
think that this special case is rather representative
since, at least in the nonrelativistic approximation,
the standard gravitational interaction operator 
$-Gm_1m_2/r$ does not act over the internal angular variables
of the two-body system. From the technical point of view
the case $s=0$ is the simplest since, as noted in the
preceding chapter, the expressions for the functions
$\Phi_n$ are defined by the characters 
of the UIR of the SU(2) group.

For quasiclassical particles the internal angular 
momentum is equal to ${\bf r}\times {\bf q}$. Since we
consider the case when the total momentum of the two-body system
and the internal angular momentum are equal to zero, we can
assume that ${\bf g}_1=(q/m_1){\bf e}_z$,  
${\bf g}_2=-(q/m_2){\bf e}_z$ and ${\bf r} =r{\bf e}_z$
where ${\bf e}_z$ is the unit vector along the $z$ axis.
Then the function $\Psi({\bf u})$ in Eq. (\ref{96}) can be
written as
\begin{eqnarray}
&&\Psi({\bf u})=exp[-\frac{i}{2}Rmu^2-\frac{m^2}{2}
(\frac{b_1^2}{m_1^2}+\frac{b_2^2}{m_2^2})(u^2+\nonumber\\
&&\frac{q^2}{m^2})+\eta u h(q,r)]   
\label{99}
\end{eqnarray}
where $u=|{\bf u}|$,
\begin{equation}
h(q,r)=mq(\frac{b_1^2}{m_1^2}+\frac{b_2^2}{m_2^2})-imr
\label{100}
\end{equation}
and $\eta$ is the cosine of the angle between ${\bf u}$
and the $z$ axis.

As noted in the preceding chapter, the functions $\Phi_n$
in $H_0$ depend only on $u$ (see Eq. (\ref{79})). 
Therefore to calculate $c_n$ in Eq. (\ref{98}) one has
first to calculate 
$$\int_{-1}^{1} exp[\eta uh(q,r)]d\eta$$
As noted in Sect. \ref{S10}, the meaning of the quantity
$b$ in Eq. (\ref{50}) is that $1/b$ is the uncertainty
of the particle velocity. Therefore $(b_1/m_1)^2$ can be
written as $1/(\Delta {\bf p}_1)^2$ where
$\Delta {\bf p}_1$ is the uncertainty of the momentum for
particle 1. Since particle 1 is quasiclassical, our choice
of the momentum variables implies that 
$muq\approx |{\bf p}_1|^2$. Therefore $mqu(b_1/m_1)^2\gg 1$
and analogously $mqu(b_2/m_2)^2\gg 1$. We conclude that
$Re(uh(r))\gg 1$. Then, as follows from Eqs. (\ref{79}),
(\ref{80}), (\ref{98}) and (\ref{99})
\begin{eqnarray}
&&c_n = \frac{const}{h(q,r)}\,\,\{\prod_{k=1}^n[\mu_1-i(2k+1)]
[\mu_2-i(2k+1)]\}\nonumber\\
&&\{\prod_{k=1}^n[w_1+(2k+1)^2][w_2+(2k+1)^2]\}^{-1/2}\nonumber\\
&&\int_{0}^{\infty}exp(-\frac{i}{2}Rmu^2-imur)A(u)u
\frac{sin(n+1)t}{sint}du
\label{101}
\end{eqnarray}
where 
\begin{equation}
A(u)=exp[-\frac{m^2}{2}(\frac{b_1^2}{m_1^2}+\frac{b_2^2}{m_2^2})
(u-\frac{q}{m})^2]
\label{102}
\end{equation}  
Since we assume that the particles
are nonrelativistic, it follows from Eq. (\ref{102}) that the 
main contribution to the integral
in Eq. (\ref{101}) is given by the region where 
$u\approx (q/m)\ll 1$ and therefore $t\approx u$.  One
might expect that Eq. (\ref{101}) is rather general in the
sense that for any reasonable choice of the quasiclassical
wave functions the expression for $A(u)$ is not necessarily 
given by Eq. (\ref{102}) but in any case $A(u)$ has a sharp
maximum around $u\approx (q/m)$.

We now write 
\begin{equation}
sin[(n+1)u]=\frac{1}{2i}(exp[i(n+1)u]-exp[-i(n+1)u])
\label{sin}
\end{equation}
and notice that the contribution of the second term to the
integral in Eq. (\ref{101}) is small since the rapidly
oscillating exponent in this expression does not have a
stationary point. At the same time the expression
corresponding to the first contribution has the stationary
point at
\begin{equation}
u_n=\frac{n+1}{Rm}-\frac{r}{R}
\label{103}
\end{equation}
Therefore, calculating the integral in Eq. (\ref{101}) by
the stationary phase method we obtain
\begin{eqnarray}
&&c_n = \frac{const}{h(r)}\{\prod_{k=1}^n[\mu_1-i(2k+1)]
[\mu_2-i(2k+1)]\}\nonumber\\
&&\{\prod_{k=1}^n[w_1+(2k+1)^2][w_2+(2k+1)^2]\}^{-1/2}\nonumber\\
&&A(u_n)exp(\frac{i}{2}mRu_n^2)
\label{104}
\end{eqnarray}

As follows from this expression, $|c_n|\approx |c_{n+1}|$.
It is also easy to see that the phases of $c_n$ and $c_{n+1}$
also are approximately the same. Indeed, when $|u_n|\ll 1$
implies that $n\ll \mu_1$ and $n\ll \mu_2$. Therefore the 
phase difference is approximately equal to
$$\frac{mR}{2}(u_{n+1}^2-u_n^2)-
2n(\frac{1}{\mu_1}+\frac{1}{\mu_2})$$ 
As follows from Eq. (\ref{103}), this quantity is indeed
negligible.

We conclude that $c_n\approx c_{n+1}$ and, as follows from
Eq. (\ref{103}), the dependence
of the coefficients $c_n$ on $n$ has a sharp maximum near
$(n/R)=q+(mr/R)$. The second term on the r.h.s. of this
equality obviously represents the correction caused by
the dS antigravity (see Sect. \ref{S6}). When $R$ is large,
this correction is small and, as a result, in the
nonrelativistic approximation we have
the situation discussed after Eq. (\ref{84}).
Therefore the mean value of the mass operator squared
in the nonrelativistic approximation is indeed given by
the standard formula (\ref{85}) assuming that
$q\ll m_1$ and $q\ll m_2$.

Let us now show that in fact
the situation discussed after Eq. (\ref{84}) takes
place in the general case, and therefore the
mean value of the mass operator squared
is precisely given by Eq. (\ref{85}). Note first
that the above derivation in the nonrelativistic case
was not quite accurate for the following reason. We 
expanded
the exponent index in powers of ${\bf u}$. However,
since the index contains a large parameter $R$, the 
terms which have been dropped are by no means small. 
Nevertheless, such a procedure is legitimate if the 
manipulations with the exponent index are
used for calculations in the stationary phase method.

\begin{sloppypar}
As argued above, the result that $\Psi({\bf u})$ is
proportional to $\Psi(0,{\bf u})$ is quite general and
therefore Eq. (\ref{98}) is correct in the general
case. It has been also shown that for the considered
class of wave functions the contribution
of the dS antigravity to the mean value of the mass
operator is small and for this reason we neglect
this contribution. 
\end{sloppypar}

In Sect. \ref{S11} it has been shown that there
exists a one-to-one correspondence between
the quantities ${\bf u}$ and ${\bf q}$ (see Eq.
(\ref{qu2})). Therefore, as follows from Eq. (\ref{13}),
a natural generalization of Eq. (\ref{101}) is
\begin{eqnarray}
&&c_n = const\,\,\{\prod_{k=1}^n[\mu_1-i(2k+1)]
[\mu_2-i(2k+1)]\}\nonumber\\
&&\{\prod_{k=1}^n[w_1+(2k+1)^2][w_2+(2k+1)^2]\}^{-1/2}\nonumber\\
&&\int_{0}^{\infty}exp[-\frac{i}{4}\mu_1 ln(m_1^2+q^2)
-\frac{i}{4}\mu_2 ln(m_2^2+q^2)]\nonumber\\
&&\frac{sin(n+1)t}{sint}A(q,r)qdq
\label{105}
\end{eqnarray}
where now $q$ is the integration variable while the
average magnitude of the relative momentum will be
denoted as $q_0$. Then one might expect that, as a
function of $q$, $A(q,r)$ has a sharp maximum in the
region where $q$ is close to $q_0$.

We will again calculate the integral in the stationary
phase method. Therefore, taking into account the definition
of $t$, we have to consider the rapidly oscillating
exponent $exp[if(q,n)]$ where
\begin{equation}
f(q,n)=-\frac{R}{2}m_1 ln(m_1^2+q^2)-
\frac{R}{2}m_2 ln(m_2^2+q^2)+(n+1)acosu_4
\label{106}
\end{equation}
A direct calculation using Eq. (\ref{qu2}) gives
\begin{equation}
\frac{\partial f(q,n)}{\partial q}=(n+1-Rq)\frac{(m_1+m_2)(m_1m_2+q^2)}
{(m_1^2+q^2)(m_2^2+q^2)}
\label{107}
\end{equation}
Therefore the stationary point is defined by
\begin{equation}
q_n=\frac{n+1}{R}
\label{108}
\end{equation}
This result is compatible with Eq. (\ref{103}) (if
the contribution of the dS antigravity is neglected)
and shows that the quantum number $n$ has the
meaning of the magnitude of the relative momentum
times $R$ not only in the nonrelativistic approximation
but exactly.

The next step is to show that $c_n$ and $c_{n+1}$ have
the same phases if $n$ and $R$ are large. As follows
from Eqs. (\ref{107}) and (\ref{108}), 
$f(q_{n+1},n)-f(q_n,n)$ is of order $1/R$ and therefore,
as follows from Eq. (\ref{105}), $c_n$ and $c_{n+1}$ 
will have approximately the same phases if
\begin{equation}
acosu_4=asin\frac{q}{(m_1^2+q^2)^{1/2}}+
asin\frac{q}{(m_2^2+q^2)^{1/2}}
\label{109}
\end{equation}
As follows from Eq. (\ref{qu2}), this expression is
indeed valid.

\section{Effect of phase difference}
\label{phase}

Before discussing the problem of interactions in
dS invariant theories, we recall well-known
facts about such a problem in Poincare invariant theory. 
It is much more complicated than in nonrelativistic 
quantum mechanics since if the Hamiltonian is interaction
dependent then some other generators are necessarily
interaction dependent too. For this reason, even for
a two-body system, the problem arises how one could 
introduce interaction into the generators without 
breaking commutation relations of the Poincare group
Lie algebra. For the first time this problem has been
solved by Bakamdjian and Thomas \cite{BT}. They showed
that if $V$ is added not to the Hamiltonian but to the mass
operator (or mass operator squared), and $V$ commutes
with the internal two-body angular momentum operator
then the required commutation relations are preserved.
As shown by several authors (see e.g. Refs. 
\cite{coester, Sokolov,rqm}), the Bakamdjian-Thomas
procedure can be extended to the case when the number
of particles is arbitrary (but fixed).

By analogy with the Bakamdjian-Thomas procedure,
it is clear from the construction of Sects. 
\ref{S11} and \ref{S12} that the mass operator
in dS invariant theory is fully defined by its
action in $H_s$. Therefore one can modify the
free dS mass operator by introducing an interaction
operator in $H_s$ commuting with the reduction of
the two-body spin operator onto $H_s$. In that case
the commutation relations (\ref{2}) will be preserved.  
However, the important
difference between Poincare and dS invariant theories
is as follows. While in the former the free and
interacting mass operators can have different spectra
(the free mass operator has only the continuous spectrum
while the interacting mass operator can also have the
discrete spectrum), in the latter they have the same
spectra and the free mass operator is not bounded
below by $m_1+m_2$. These facts have been discussed in 
Chap. \ref{Ch2} and it has been noted that such features
pose the problem whether the very notion of interaction
is needed at all. 

As it has been shown in the preceding section, if
the two-body wave function is the product of the
single-particle wave functions having the form (\ref{49}),
then the mean value of the mass operator
is given by Eq. (\ref{85}) when $R$ is large.
Although this fact could be expected, the above 
derivation shows 
that if the Poincare invariant theory is treated as 
a special case of the dS invariant one in the limit 
when $R$ is large, then
rather delicate cancellations are required to ensure the
condition $c_n\approx c_{n+1}$.

We again consider a quasiclassical two-body system
such that the internal wave function in momentum space
has a sharp maximum at $q=q_0$ and the internal 
wave function in coordinate space has a sharp maximum 
at $r=r_0$. Since each space $H_s$ has the basis
characterized by the functions $\Psi_n$, the projection
of each wave function $\Psi$ onto $H_s$ is fully 
characterized by the coefficients $c_n=(\Psi_n,\Psi)$.
Let $c_n^{(0)}$ be the coefficients calculated in the
preceding section. 

Consider a wave function defined as
follows. Its projection onto $H_s$ is characterized by
the coefficients $c_n=c_n^{(0)}exp[i\eta_n(r_0)]$
where $\eta_n(r_0)$ are real functions such that
$\eta_{n+1}(r_0)-\eta_n(r_0)=\pm \delta(r_0)$ and
$\delta(r_0)$ is given by Eq. (\ref{90}). Since
$c_{n+1}^{(0)}\approx c_n^{(0)}$, we have
precisely the situation discussed in Sect. 
\ref{relativistic} when the mean value of the mass
operator squared is given by Eq. (\ref{89}). 

The class of wave functions for which the above
property is satisfied is rather wide. For example,
a possible choice of $\eta_n(r_0)$ is 
$\eta_n(r_0)=\pm n\delta_0(r_0)$. Since the 
coefficients $c_n$ are defined up to an
arbitrary overall phase factor, an interesting
scenario is such that $\eta_n(r_0)=0$ if $n$ is even
and $\eta_n(r_0)=\delta_0(r_0)$ if $n$ is odd.
In that case the phase difference between $c_n$
and $c_n^{(0)}$ does not exceed $\delta_0(r_0)$ for
all values of $n$.

The problem arises whether the wave functions with 
such coefficients $c_n$ are compatible
with the fact that the two-body system is 
quasiclassical. Since there exist many scenarios
when the result (\ref{89}) takes place, this problem
seems to be rather complicated. In the subsequent 
section we consider a hypothesis that a natural
explnation of the phase difference between $c_n$
and $c_n^{(0)}$ can be obtained in quantum theory 
over a Galois field.
  
In the framework of General Relativity, Newtonian
gravity is the first approximation valid when the
particles are nonrelativistic and $\delta_0(r_0)\ll 1$.
The internal classical two-body Hamiltonian taking
into account the corrections of order $(q/m)^2$ and
$\delta_0(r)^2$ is given by (see e.g. problem 3 in
section 106 of Ref. \cite{LL2})      
\begin{eqnarray}
&&M = m_1+m_2 + \frac{q^2}{2m}-\frac{Gm_1m_2}{r}-
\frac{q^4}{8}(\frac{1}{m_1^3}+\frac{1}{m_2^3})-
\frac{G}{2r}[3q^2(\frac{m_1}{m_2}+\nonumber\\
&&\frac{m_2}{m_1})+
7q^2+\frac{({\bf q}{\bf r})^2}{r^2}]+
\frac{G^2}{2r^2}m_1m_2(m_1+m_2)
\label{116}
\end{eqnarray}
In quasiclassical approximation the angular momentum
is equal to ${\bf r}\times {\bf q}$ and therefore 
$q^2r^2-({\bf q}{\bf r})^2\approx {\bf j}^2$ where
${\bf j}$ is the quasiclassical internal angular
momentum. Therefore with the same accuracy 
\begin{eqnarray}
&&M^2 = M_0^2-\frac{2Gm_1m_2(m_1+m_2)}{r}+
\frac{G^2}{r^2}m_1m_2[(m_1+m_2)^2+\nonumber\\
&&m_1m_2]-\frac{3G(m_1+m_2)}{m_1m_2r}q^2[(m_1+m_2)^2+
m_1m_2]+\nonumber\\
&&\frac{G}{r^3}{\bf j}^2(m_1+m_2)
\label{117}
\end{eqnarray}
where $M_0^2$ is given by Eq. (\ref{85}).

In quantum theory Eq. (\ref{117}) should be treated as
the mean value of the mass operator squared, and $q$ and
$r$ should be treated as the mean values of the corresponding
quantities, i.e. $q$ should be replaced by $q_0$ and $r$ 
by $r_0$. Since we consider the mass distribution in
each space $H_s$ separately, ${\bf j}^2$ should be
replaced by the eigenvalue of the internal angular
momentum squared in $H_s$, i.e. $j(j+1)$. 

By analogy with the nonrelativistic case, one can
show that the result compatible with Eq. (\ref{117}) 
can be obtained for wave functions characterized by the
coefficients $c_n=c_n^{(0)}exp[i\eta_n(r_0,q_0)]$
such that $\eta_{n+1}(r_0,q_0)-\eta_n(r_0,q_0)=
\pm \delta(r_0,q_0)$,
\begin{eqnarray}
\delta(r_0,q_0)=\delta_0(r_0)[1-a_1\delta_0(r_0)^2-
\frac{j(j+1)}{8m_1m_2r_0^2}+a_2q_0^2]
\label{118}
\end{eqnarray} 
and the coefficients $a_1$ and $a_2$ are given by 
\begin{equation}
a_1=\frac{1}{12}+\frac{m_1m_2}{8(m_1+m_2)^2}\quad
a_2=\frac{1}{2m_1^2}+\frac{1}{2m_2^2}+\frac{9}{4m_1m_2}
\label{119}
\end{equation}
(note that in the nonrelativistic approximation
$\delta(r_0,q_0)=\delta_0(r_0)$).
 
\chapter{Quantum theory over a Galois field}
\label{Ch5}

\section{What mathematics is most suitable for quantum 
physics?}
\label{SG1}

In the preceding chapters it has been shown that the
dS invariant quantum theory has unusual features which 
have no analogs in Poincare
and AdS invariant theories. As a consequence of the
fact that the spectrum of the dS mass operator is not
bounded below by the value of $m_1+m_2$, for a class
of two-body wave functions, the mean value of the 
mass operator is compatible with Newtonian
gravity and post-Newtonian corrections. This poses
a very interesting problem, whether gravity can
be explained without using the notion of interaction
at all. However, such wave functions contain 
additional phase factors and it is not clear whether
they are compatible with our understanding of the position
operator in quantum theory. Another problem in the
standard quantum theory is that the notion of 
interaction depends on the choice of the form 
of the single-particle generators. 

In the present chapter we argue that quantum theory 
over a Galois field is more natural than 
the standard quantum theory based on complex numbers.
We believe that some properties of the GFQT give
indications that gravity can be explained in the
framework of this theory. Since the absolute
majority of physicists are not familiar with
Galois fields, our first goal in this chapter is
to convince the reader that the notion of Galois
fields is not only very simple and elegant, but
also is a natural basis for quantum physics. We
will follow mainly our arguments given in Ref.
\cite{hep2}. If a reader wishes to learn Galois
fields on a more fundamental level, he or she
might start with standard textbooks (see e.g.
Ref. \cite{VDW}).

The existing quantum theory is based on 
standard mathematics containing the notions of
infinitely small and infinitely large.

The notion of infinitely small is based on our
everyday experience that any macroscopic object can be 
divided by two, three and even a million
parts. But is it possible to divide by two or three 
the electron or neutrino? It seems obvious that 
the very existence of elementary particles 
indicates that the standard division has only a 
limited sense. Indeed, let us consider, for example,
the gram-molecule of water having the mass 18 grams. 
It contains the Avogadro number of molecules 
$6\cdot 10^{23}$. We can divide this 
gram-molecule by ten, million, billion, but when we 
begin to divide by numbers greater than the Avogadro 
one, the division operation loses its sense. 

The notion of infinitely large is based on our belief
that {\it in principle} we can operate with any
large numbers. Suppose we wish to verify experimentally 
whether addition is commutative, i.e. whether a + b = 
b + a is always satisfied. If our Universe is finite 
and contains not more than N elementary particles then 
we shall not be able to do this if $a + b > N$. 
In particular, if the Universe is finite then it is
impossible in principle to build a computer operating
with any large number of bits.

It is interesting to note that the number of 
elementary particles in the Universe might be not
so immense as one could think. Indeed, if the radius
of the Universe is 20 billion light years, and the
average density of matter is $10^{-29}g/cm^3$ then
the mass of the Universe is of order $10^{57}g$. If
we assume that the average energy of elementary
particles is $1eV$ then the number of elementary
particles in the Universe is "only" of order $10^{90}$.
Note, for example, that the largest known prime
number has 4033946 digits \cite{primes}.

As noted in Sect. \ref{S5}, the dS and Poincare
masses are related to each other as $\mu=2Rm$.
In our units $m=2/l_C$ where $l_C$ is the particle
Compton wave length. Thefore $\mu$ is roughly
the ratio of the dS radius to the Compton wave
length. Hence even the dS masses of elementary
particles are very large. For example,
if $R$ is, say, 20 billion light years then  
the dS mass of the electron is of order
$10^{39}$. My observation is that
physicists are usually surprised by this fact
and their first reaction is that the dS mass
is something unrealistic. However, the dS mass
is dimensionless and therefore more fundamental
than the Poincare mass which depends on the
system of units. The electron dS mass might be
an indication that the electron is not a truly
elementary particle, but at present such an
assumption is highly speculative. 

As shown in Sect. \ref{S4}, the UIRs defined
by Eqs. (\ref{9}) and (\ref{12}) are related to
each other by the unitary transformation (\ref{13}).
Note that the exponent index in Eq. (\ref{13}) is
very large. For example, if $R$ is again taken to
be 20 billion light years then the dS mass of the
Earth is of order $10^{94}$. If the exponent is
defined in a standard way by the Taylor series
then no existing computer can compute the exponent
with such a large index. One can notice
that in fact the index in Eq. (\ref{13}) should
be taken modulo $2\pi$, but in that case the
result will be sensitive to the accuracy 
$\pi$ is known with. Suppose, however, that this
problem is solved. Then we are facing another
serious difficulty. The  
quasiclassical approximation implies that all
the integrals in question can be calculated by
the stationary phase method. It is well known
that the idea of the method is
to perturb the contour of integration in such a
way that in some vicinity of the stationary point
$x_0$ the integral can be approximated by
$$\int_{-\infty}^{\infty} exp[-b(x-x_0)^2]dx 
\quad (a> 0).$$
In the case of Eq. (\ref{101}) the role of $x$
is played by the nonrelativistic velocity and
$b$ is the dS mass. Therefore the integral in
Eq. (\ref{101}) is mainly defined by a 
very narrow region of velocities having the
width of order $10^{-47}c$. It is difficult to
imagine that such an accuracy is meaningful. 

Many physicists believe that (in contrast with
the Dirac hypothesis \cite{Dirhyp}) the 
gravitational constant $G$ is fundamental
and therefore the Planck mass is fundamental
too. The electron Planck mass is very small
(of order $10^{-23}$) in agreement with the
assumption that gravity is unified with the
other interactions at the Planck scale 
$10^{-5}g$. On the other hand, as noted in
Sect. \ref{S3}, in quantum dS (and AdS) 
invariant theories there is no place for
quantities having the dimension $(length)^2$
in units $\hbar/2 = c=1$. 

Let us look at mathematics from the point of 
view of the famous Kronecker expression: 'God made 
the natural numbers, all else is the work of man'.  
Indeed, the natural numbers 0, 1, 2... 
(we treat zero as a natural number) have a 
clear physical meaning. However only two operations
are always possible in the set of natural 
numbers: addition and multiplication. 

In order to 
make addition reversible, we introduce negative 
integers -1, -2 etc. Then, instead of the set of 
natural numbers we can work
with the ring of integers where three operations 
are always possible:
addition, subtraction and multiplication. However, 
the negative numbers do not have a direct physical 
meaning (we cannot say, for example, 'I have 
minus two apples'). Their only role is to make 
addition reversible. 

The next step is the transition to the field of 
rational numbers in which all
four operations excepting division by zero are possible.
However, as noted above, division has only a limited
sense. 

In mathematics the notion of linear space is 
widely used, and such important
notions as the basis and dimension are meaningful only
if the space is considered over a field or body. 
Therefore if we start from natural numbers and wish 
to have a field, then we have to introduce negative
and rational numbers. However, if, instead of all 
natural numbers, we consider
only $p$ numbers 0, 1, 2, ... $p-1$ where $p$ is 
prime, then we can easily construct a field without 
adding any new elements. This construction, called
Galois field, contains nothing that could prevent 
its understanding even by pupils of elementary 
schools.  

Let us denote the set of numbers 0, 1, 2,...$p-1$ 
as $F_p$. Define addition and multiplication as usual 
but take the final result modulo $p$. For simplicity,
let us consider the case $p=5$. Then $F_5$ is the set 0, 
1, 2, 3, 4. Then
$1+2=3$ and $1+3=4$ as usual, but $2+3=0$, $3+4=2$ etc. 
Analogously, $1\cdot 2=2$,
$2\cdot 2=4$, but $2\cdot 3 =1$, $3\cdot 4=2$ etc. 
By definition, the element
$y\in F_p$ is called opposite to $x\in F_p$ and is 
denoted as $-x$ if $x+y=0$ in
$F_p$. For example, in $F_5$ we have -2=3, -4=1 etc. 
Analogously $y\in F_p$ is
called inverse to $x\in F_p$ and is denoted as 
$1/x$ if $xy=1$ in $F_p$. 
For example, in $F_5$ we have 1/2=3, 1/4=4 etc. It is 
easy to see that
addition is reversible for any natural $p>0$ but for 
making multiplication
reversible we should choose $p$ to be a prime. 
Otherwise the product of two
nonzero elements may be zero modulo $p$. If $p$ is 
chosen to be a prime then
indeed $F_p$ becomes a field without introducing any 
new objects (like negative numbers as fractions). For 
example, in this field each element can obviously be 
treated as positive and negative {\bf simultaneously}! 

One might say: well, this is beautiful but impractical 
since in physics and 
everyday life 2+3 is always 5 but not 0. Let us suppose, 
however that fundamental
physics is described not by 'usual mathematics' but by 
'mathematics modulo $p$' 
where $p$ is a very large number.
Then, operating with numbers much smaller than $p$ we 
shall not notice this $p$,
at least if we only add and multiply. We will feel a 
difference between 'usual mathematics' and 'mathematics 
modulo p' only while operating with numbers
comparable with $p$. 

We can easily extend the correspondence between $F_p$ and 
the ring of integers $Z$
in such a way that subtraction will also be included. 
To make it clearer we note
the following. Since the field $F_p$ is cyclic (adding 
1 successively, we will
obtain 0 eventually), it is convenient to visually depict 
its elements by the 
points of a circle of the radius $p/2\pi$ on the plane $(x,y)$. 
In Fig. 1.1 only a
part of the circle near the origin is depicted. 
\begin{figure}[!ht]
\centerline{\scalebox{1.1}{\includegraphics{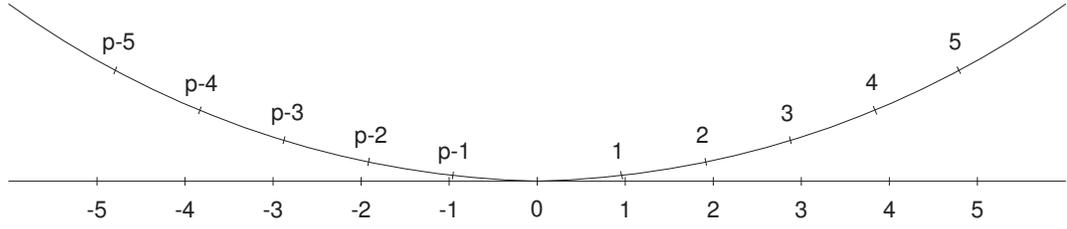}}}%
\caption{%
  Relation between $F_p$ and the ring of integers%
}%
\label{Fig.1}
\end{figure}
Then the distance between
neighboring elements of the field is equal to unity, and 
the elements 
0, 1, 2,... are situated on the circle counterclockwise. At 
the same time we depict the elements of $Z$ as usual 
such that each element $z\in Z$ is depicted by 
a point with the
coordinates $(z,0)$. We can denote the elements of $F_p$ 
not only as 0, 1,... $p-1$
but also as 0, $\pm 1$, $\pm 2,$,...$\pm (p-1)/2$, and such 
a set is called the
set of minimal residues. Let $f$ be a map from $F_p$ to Z, 
such that the element
$f(a) \in Z$ corresponding to the minimal residue $a$ has 
the same notation as
$a$ but is considered as the element of $Z$. Denote 
$C(p) = p^{1/(lnp)^{1/2}}$ and
let $U_0$ be the set of elements $a\in F_p$ such that 
$|f(a)|<C(p)$. Then if
$a_1,a_2,...a_n\in U_0$ and $n_1,n_2$ are such natural 
numbers that
\begin{equation}
n_1<(p-1)/2C(p),\,\,n_2<ln((p-1)/2)/(lnp)^{1/2}
\end{equation}
\label{G1}
then 
$$f(a_1\pm a_2\pm...a_n)=f(a_1)\pm f(a_2)\pm ...f(a_n)$$ 
if $n\leq n_1$ and
$$f(a_1 a_2...a_n)=f(a_1)f(a_2) ...f(a_n)$$ if $n\leq n_2$.
Thus though $f$ is not a homomorphism of rings $F_p$ and $Z$, 
but if $p$ is sufficiently large, then for a sufficiently 
large number of elements of $U_0$ the
addition, subtraction and multiplication are 
performed according to the same rules
as for elements $z\in Z$ such that $|z|<C(p)$. 
Therefore $f$ can be regarded as a 
local isomorphism of rings $F_p$ and $Z$.

The above discussion has a well known historical 
analogy. For many years people believed
that our Earth was flat and infinite, and only 
after a long period of time they realized that 
it was finite and had a curvature. 
It is difficult to notice the curvature when we 
deal only with 
distances much less than the radius of the
curvature $R$. Analogously one might think that the set of 
numbers describing physics has a curvature defined 
by a very large 
number $p$ but we do not notice it when
we deal only with numbers much less than $p$.

Our arguments imply that the standard field of 
real numbers will not be fundamental in future 
quantum physics (although
there is no doubt that it is relevant for macroscopic
physics). Let us discuss this question in a greater detail
(see also Ref. \cite{FJD1}).
The notion of real numbers can be fundamental only if the
following property is valid: for any $\epsilon>0$, it is
possible (at least in principle) to build a computer
operating with a number of bits $N(\epsilon)$ such that
computations can be performed with the accuracy better
than  $\epsilon$. It is clear that this is not the case 
if, for example, the Universe is finite. 
 
Let us note that even for elements from $U_0$ the result of 
division in the field
$F_p$ differs generally speaking, from the corresponding 
result in the field of rational number $Q$. For example the 
element 1/2 in $F_p$ is a very large number
$(p+1)/2$. For this reason one might think that physics based 
on Galois
fields has nothing to with the reality. We will see in 
the subsequent section that this is not so since the 
spaces describing quantum systems are projective.

By analogy with the field of complex numbers, we can 
consider a set $F_{p^2}$
of $p^2$ elements $a+bi$ where $a,b\in F_p$ and $i$ is 
a formal element such
that $i^2=1$. The question arises whether $F_{p^2}$ is a 
field, i.e.
we can define all the four operations excepting 
division by zero.
The definition of addition, subtraction and multiplication 
in $F_{p^2}$
is obvious and, by analogy with the field of complex 
numbers, we can
try to define division as $1/(a+bi)\,=a/(a^2+b^2)\,-ib/(a^2+b^2)$.
This definition can be meaningful only if $a^2+b^2\neq 0$ in $F_p$
for any $a,b\in F_p$ i.e. $a^2+b^2$ is not divisible by $p$.
Therefore the definition is meaningful only if $p$ {\it cannot}
be represented as a sum of two squares and is meaningless otherwise.
We will not consider the case $p=2$ and therefore $p$ is necessarily odd.
Then we have two possibilities: the value of $p\,(mod \,4)$ is either 1
or 3. The well known result of the number theory (see e.g. the 
textbooks \cite{VDW}) is that a prime number $p$ can be 
represented as a sum of two squares only in the former case
and cannot in the latter one. Therefore the above construction of
the field $F_{p^2}$ is correct only if $p\,(mod \,4)\,=\,3$.
The first impression is that if Galois fields can somehow 
replace the conventional field of complex numbers then this 
can be done only for $p$ satisfying $p\,(mod \,4)\,=\,3$ and
therefore the case $p\,(mod \,4)\,=\,1$ is of no interest for
this purpose. We will see in the subsequent section that 
correspondence between complex numbers and Galois fields
containing $p^2$ elements can also be established if
$p\,(mod \,4)\,=\,1$. In general, it is well known (see e.g.
Ref. \cite{VDW}) that any Galois field consists of $p^n$
elements where $p$ is prime and $n>0$ is natural. The numbers
$p$ and $n$ define the field $F_{p^n}$ uniquely up to 
isomorphism and $p$ is called the characteristic of the
Galois field.

\section{Modular representations of Lie algebras}
\label{SG2}

A well-known historical fact is that quantum mechanics
has been originally proposed by Heisenberg and 
Schroedinger in two forms which seemed fully
incompatible with each other. While in the Heisenberg matrix
formulation quantum states are described by infinite
columns and operators --- by infinite matrices, in the
Schroedinger wave formulations the states are described
by functions and operators --- by differential operators.
It has been shown later by von Neumann and others that
the both formulations are mathematically equivalent. 

Quantum theory over a Galois field (GFQT) can be treated
as a version of the matrix formulation when complex numbers 
are replaced by elements of a Galois field. We will see below
that in that case the columns and matrices are automatically
truncated in a certain way, and therefore the theory becomes
finite-dimensional (and even finite since any Galois field
is finite).

In conventional quantum theory the state of a system is 
described by a vector $\tilde x$ from a separable Hilbert 
space $H$. We will use a "tilde"  to denote elements
of Hilbert spaces and complex numbers while elements
of linear spaces over Galois fields and elements of the
fields will be denoted without a "tilde". 

Let $(\tilde e_1,\tilde e_2,...)$ be a basis in $H$. This
means that $\tilde x$ can be represented as
\begin{equation}
\tilde x =\tilde c_1 \tilde e_1+\tilde c_2 \tilde e_2+...
\label{G2}
\end{equation}
where $(\tilde c_1,\tilde c_2,...)$ are complex numbers. 
It is assumed
that there exists a complete set of commuting selfadjoint 
operators $(\tilde A_1,\tilde A_2,...)$ in $H$ such that
each $\tilde e_i$ is the eigenvector of all these operators:
$\tilde A_j\tilde e_i =\tilde \lambda_{ji}\tilde e_i$. Then the
elements $(\tilde e_1,\tilde e_2,...)$ are mutually orthogonal:
$(\tilde e_i,\tilde e_j)=0$ if $i\neq j$ where (...,...) is
the scalar product in $H$. In that case the coefficients can
be calculated as
\begin{equation}
\tilde c_i = \frac{(\tilde e_i,\tilde x)}{(\tilde e_i,\tilde e_i)}
\label{G3}
\end{equation}
Their meaning is that 
$|\tilde c_i|^2(\tilde e_i,\tilde e_i)/(\tilde x,\tilde x)$
represents the probability to find $\tilde x$ in the state
$\tilde e_i$. In particular, when $\tilde x$ and the basis
elements are normalized to one, the probability is exactly
equal to $|\tilde c_i|^2$.

Let us note first that the Hilbert space contains a big
redundancy of elements, and we do not need to know all of
them. Indeed, with any desired accuracy we can approximate
each $\tilde x\in H$ by a finite lnear combination
\begin{equation}
\tilde x =\tilde c_1 \tilde e_1+\tilde c_2 \tilde e_2+...
\tilde c_n\tilde e_n 
\label{G4}
\end{equation}
where $(\tilde c_1,\tilde c_2,...\tilde c_n)$ are
rational complex numbers. In turn, the set Eq. (\ref{G4})
is redundant too. Indeed, we can use the fact that Hilbert
spaces in quantum theory are projective: $\psi$ and 
$c\psi$ represent the same physical state. Then we can
multiply both parts of Eq. (\ref{G4}) by a common
denominator of the numbers 
$(\tilde c_1,\tilde c_2,...\tilde c_n)$. As a result, we
can always assume that in Eq. (\ref{G4}) 
$\tilde c_j=\tilde a_j +i\tilde b_j$ where $\tilde a_j$
and $\tilde b_j$ are integers. If it is convenient, we can
approximate all physical states by another sets of elements.
For example, we can use Eq. (\ref{G4}), where
$\tilde c_j=\tilde a_j +\sqrt{2}i\tilde b_j$ and
$\tilde a_j$ and $\tilde b_j$ are integers. 

The meaning of the fact that Hilbert
spaces in quantum theory are projective is very clear.
The matter is that not the probability itself but the
relative probabilities of different measurement outcomes
have a physical meaning. 
We believe, the notion of probability is a
good illustration of the Kronecker expression about
natural numbers (see Sect. \ref{SG1}). Indeed, this notion 
arises as follows. Suppose that conducting 
experiment $n$ times we have seen the first event $n_1$ times, 
the second event $n_2$ times etc. such that $n_1 + n_2 + ... = n$. 
We introduce the quantities $w_i(n) = n_i/n$ (these quantities 
depend on $n$) and $w_i = lim\, w_i(n)$ when $n \rightarrow 
\infty$. Then $w_i$ is called the probability of the 
$ith$ event. We see that all the information about the 
experiment is given by a set of natural numbers. 
However, in order to define 
probabilities, people introduce additionally the notion of 
rational numbers and the notion of limit. Of course,  
the standard notion of probability can be used even if quantum 
theory is based entirely on natural numbers, but one should 
realize that this is only a convention on how to describe 
the measurement outcomes.

The Hilbert space is an example of a linear space over the
field of complex numbers. Roughly speaking this means that
one can multiply the elements of the space by the elements 
of the field and use the properties
$\tilde a(\tilde b\tilde x)=(\tilde a\tilde b)\tilde x$
and $\tilde a(\tilde b\tilde x+\tilde c\tilde y)= 
\tilde a\tilde b\tilde x +\tilde a\tilde c\tilde y$ where
$\tilde a,\tilde b,\tilde c$ are complex numbers and
$\tilde x,\tilde y$ are elements of the space. The fact that 
complex numbers form a field is important for such notions
as linear dependence and the dimension of spaces over complex
numbers. 

In general, it is possible to consider linear
spaces over any fields (see any textbook on modern
algebra, e.g. \cite{VDW}). It is natural to assume that
in the GFQT the states of physical
systems should be described by elements of linear spaces over
a Galois field. Since we wish to have a correspondence
with the conventional theory, we assume that the Galois field
in question contains $p^2$ elements where $p$ is the
characteristic of the field. In Sect. \ref{SG1} we discussed
the correspondence between the ring of integers and
the field $F_p$. It has been also noted that if 
$p=3\,(mod\,4)$ then the elements of $F_{p^2}$ can be
written as $a+bi$ where $a,b\in F_p$. We will now discuss
the construction of the field $F_{p^2}$ in more general
cases.

The field $F_{p^2}$ can be constructed by means of the 
standard extension of the field $F_p$. Let the equation
$\kappa^2=-a_0$ $(a_0\in F_p)$ have no solutions in $F_p$.
Then $F_{p^2}$ can be formally described as a set of
elements $a+b\kappa$, where $a,b\in F_p$ and $\kappa$ 
satisfies the condition $\kappa^2=-a_0$. The actions in
$F_{p^2}$ are defined in the natural way. The condition
that the equation $\kappa^2=-a_0$ has no solutions in 
$F_p$ is important in order to ensure that any nonzero 
element from $F_{p^2}$ has an inverse.
Indeed, the definition 
$(a+b\kappa)^{-1}=(a-b\kappa)/(a^2+b^2a_0)$ is correct 
since the denominator can be equal to zero only if both,
$a$ and $b$ are equal to zero.

Consider three cases.

(1) $a_0=1$. In that case it is natural to write $i$
instead of $\kappa$. We must be sure that the element -1
in $F_p$ cannot be represented as a square of another
element from $F_p$. In the number theory this is
expressed by saying that -1 is not a quadratic residue
in $F_p$. It is well known (see e.g. Ref. \cite{VDW})
that this is the case when $p=3\,(mod\,4)$ and is not
the case when $p=1\,(mod\,4)$. This fact is closely 
related to one mentioned in Sect. \ref{SG1} that a
prime $p$ can be represented as a sum of two squares 
only if $p=1\,(mod\,4)$. 

(2) $a_0=2$. Assume that $p=1\,(mod\,4)$. Since -1 is
a quadratic residue in $F_p$ in this case, then the problem
is that 2 should not be a quadratic residue in this field.
This is the case (see e.g. Ref. \cite{VDW}) 
if $p=5\,(mod\,8)$.

(3) $a_0=3$. Again, assume that $p=1\,(mod\,4)$. 
Then 3 should not be a quadratic residue in $F_p$. 
This is, for example, the case (see e.g. Ref. 
\cite{VDW}),
when $p$ is the prime of Fermat's type, i.e. $p=2^n+1$.

In the general case, the field $F_p$ can be extended to a field
containing $p^2$ elements as follows. First we note a well known
fact \cite{VDW} that $F_p$ is a cyclic group with respect
to multiplication. There exists at least one element $r\in F_p$
such that $(r,r^2,...r^{p-1}=1)$ represent the set $(1,2,...p-1)$
in some order. The element $r$ is called the primitive root of
the field $F_p$. It is clear from this observation that $F_p$ 
contains the equal number $(p-1)/2$ of quadratic residues and
nonresidues: the elements represented as odd powers of $r$ are
nonresidues while those represented by even powers of $r$ ---
residues. In particular, $r$ is not a quadratic residue. Now
we can formally introduce $\kappa$ as an element satisfying
the condition $\kappa^2=r$, and $F_{p^2}$ as a set of
elements $a+b\kappa$ $(a,b\in F_p)$. In this case the
definition
$(a+b\kappa)^{-1}=(a-b\kappa)/(a^2-b^2r)$ is correct since
$a^2-b^2r\neq 0$ if $a\neq 0$ or $b\neq 0$.

The above observation shows that if $F_p$ is extended to
$F_{p^2}$ then any element of $F_p$ has a square root
belonging to $F_{p^2}$. 

It is well known (see e.g. Ref. \cite{VDW}) that the 
field of $p^2$ elements has only one nontrivial
automorphism. In the above cases it can be defined as
$a+b\kappa\rightarrow a-b\kappa$ or as 
$a+b\kappa\rightarrow (a+b\kappa)^p$. We will use a bar
to denote this automorphism. This means that if
$c=a+b\kappa$ $(a,b\in F_p)$ then $\bar{c}=a-b\kappa$.

By analogy with conventional quantum theory, we
require that linear spaces V over $F_{p^2}$, used for
describing physical states, are supplied by a
scalar product (...,...) such that for any $x,y\in V$
and $a\in F_{p^2}$, $(x,y)$ is an element of $F_{p^2}$
and the following properties are satisfied: 
\begin{equation}
(x,y) =\overline{(y,x)},\quad (ax,y)=\bar{a}(x,y),\quad 
(x,ay)=a(x,y)
\label{G5}
\end{equation}  

In Sect. \ref{SG1} we discussed a map $f$ from $F_p$
to the ring of integers $Z$. We can extend the map in such
a way that it maps $F_{p^2}$ to a subset of complex numbers.
For example, if $p=3\,(mod\,4)$, $c\in F_{p^2}$, $c=a+bi$
$(a,b\in F_p)$ then we can define $f(c)$ as $f(a)+f(b)i$.
If $a_0=-2$ and $c=a+b\kappa$, we define 
$f(c)=f(a)+\sqrt{2}f(b)i$. Analogously, if $a_0=-3$ and 
$c=a+b\kappa$, we define $f(c)=f(a)+\sqrt{3}f(b)i$.
Let $U$ be a subset of elements $c=a+b\kappa$ where 
$a,b\in U_0$ (see Sect. \ref{SG1}). Then, for the elements
from $U$, addition, subtraction and multiplication look 
the same as for the
elements $f(c)$ of the corresponding subset in the field of 
complex numbers. This subset is of the form $\tilde a +
\tilde b i$ in the first case, $\tilde a + \sqrt{2}
\tilde b i$ in the second case and $\tilde a + \sqrt{3}
\tilde b i$ in the third one. Here $\tilde a, \tilde b \in Z$.    

We will always consider only finite dimensional spaces $V$ over
$F_{p^2}$. Let $(e_1,e_2,...e_N)$ be a basis in such a space. 
Consider subsets in $V$ of the form $x=c_1e_1+c_2e_2+...c_ne_n$ 
where for any $i,j$
\begin{equation}
 c_i\in U,\quad (e_i,e_j)\in U
\label{G6}
\end{equation}
On the other hand, as noted above, in conventional quantum
theory we can describe quantum states by subsets of the form   
Eq. (\ref{G4}). If $n$ is much less than $p$,
\begin{equation}
f(c_i)=\tilde c_i,\quad f((e_i,e_j))=(\tilde e_i,\tilde e_j)
\label{G7}
\end{equation}
then we have the correspondence between the description of
physical states in projective spaces over $F_{p^2}$ on the
one hand and projective Hilbert spaces on the other. This means
that if $p$ is very large, then for a large number 
of elements from $V$,
linear combinations with the coefficients belonging to $U$ and
scalar products look in the same way as for the elements from
a corresponding subset in the Hilbert space. 

In the general case a scalar product in $V$ does not define
any positive definite metric, and thus there is no
probabilistic interpretation for all the elements from $V$. 
In particular, $(e,e)=0$ does not necessarily imply that $e=0$. 
However, the probabilistic interpretation exists for such a
subset in $V$ that the conditions (\ref{G7}) are satisfied.
Roughly speaking this means that for elements $c_1e_1+...c_ne_n$
such that $(e_i,e_i),c_i{\bar c}_i\ll p$, $f((e_i,e_i))>0$
and $c_i{\bar c}_i>0$ for all $i=1,...n$, the 
probabilistic interpretation is valid. 
It is also possible to explicitly construct 
a basis $(e_1,...e_N)$ such that $(e_j,e_k)=0$ for $j\neq k$ and
$(e_j,e_j)\neq 0$ for all $j$ (see the subsequent section). 
Then $x=c_1e_1+...c_Ne_N$ $(c_j\in
F_{p^2})$ and the coefficients are uniquely defined by 
$c_j=(e_j,x)/(e_j,e_j)$. 

 As usual, if $A_1$ and $A_2$ are linear operators in $V$ such that
\begin{equation}
(A_1x,y)=(x,A_2y)\quad \forall x,y\in V
\label{G8}
\end{equation}
they are said to be conjugated: $A_2=A_1^*$. It is easy to see
that $A_1^{**}=A_1$ and thus $A_2^*=A_1$. If $A=A^*$ then the
operator $A$ is said to be Hermitian.

 If $(e,e)\neq 0$, $Ae=ae$, $a\in F_{p^2}$, and $A^*=A$, then it
is obvious that $a\in F_p$. At the same time, there exist 
possibilities (see e.g. Ref. \cite{lev4}) when $e\neq 0$, 
$(e,e)=0$ and the element $a$ is imaginary, i.e. $a=b\kappa$, 
$b\in F_p$. Further, if
\begin{eqnarray}
&&Ae_1=a_1e_1,\quad Ae_2=a_2e_2,\quad (e_1,e_1)\neq 0,\nonumber\\
&& (e_2,e_2)\neq 0, \quad a_1\neq a_2
\label{G9}
\end{eqnarray}
then as in the usual case, one has $(e_1,e_2)=0$. At the same time,
the situation when
\begin{eqnarray}
&&(e_1,e_1)=(e_2,e_2)=0,\quad (e_1,e_2)\neq 0,\nonumber\\
&& a_1=b_1\kappa, \quad a_2=b_2\kappa\quad (b_1,b_2\in F_p)
\label{G10}
\end{eqnarray}
is also possible \cite{lev4}.

Let now $(A_1,...A_k)$ be a set of Hermitian commuting operators
in $V$, and $(e_1,...e_N)$ be a basis in $V$ with the  properties
described above, such that $A_je_i=\lambda_{jl}e_i$. Further, let
$({\tilde A}_1,...{\tilde A}_k)$ be a set of Hermitian  commuting
operators  in  some  Hilbert  space  $H$,   and
$(\tilde e_1,\tilde e_2,...)$ be some basis in $H$ such that
$\tilde A_je_i={\tilde \lambda}_{ji}\tilde e_i$.
Consider a subset $c_1e_1+c_2e_2+...c_ne_n$ in $V$ such that,
in addition to the conditions (\ref{G7}), the elements $e_i$
are the eigenvectors of the operators $A_j$ with $\lambda_{ji}$
belonging to $U$ and such that 
$f(\lambda_{ji})={\tilde \lambda}_{ji}$. Then the action of the 
operators on such
elements have the same form as the action of corresponding 
operators on the subsets of elements in Hilbert spaces discussed
above. 

Summarizing this discussion, we conclude that if $p$ is large then
there exists a correspondence between the description of physical
states on the language of Hilbert spaces and selfadjoint operators
in them on the one hand, and on the language of linear spaces over
$F_{p^2}$ and Hermitian operators in them on the other.

The field of complex numbers is algebraically closed (see 
standard textbooks on modern algebra, e.g. Ref. \cite{VDW}).
This implies that any equation of the $nth$ order in this field
always has $n$ solutions. This is not, generally speaking, the
case for the field $F_{p^2}$. As a consequence, not every linear
operator in the finite-dimensional space over $F_{p^2}$ has an 
eigenvector (because the characteristic equation may have no 
solution in this field). One can define a field of
characteristic $p$ which is algebraically closed and contains
$F_{p^2}$. However such a field will necessarily be infinite and 
we will not use it. For the same reason we will not discuss
theories based on $p$-adic or adelic fields (see e.g.
Ref. \cite{Volovich}) although they have several interesting
features (for example, in the adelic case there is
no distinguished value of $p$). We will see in the 
subsequent section
that uncloseness of the field $F_{p^2}$ does not prevent one
from constructing physically meaningful representations
describing elementary particles in the GFQT.

In physics one usually considers Lie  algebras over $R$ and
their representations by Hermitian  operators in
Hilbert spaces. It is clear that the analogs of such
representations in our case are the representations of Lie
algebras over $F_p$  by Hermitian operators 
in spaces over $F_{p^2}$.
Representations  in  spaces  over  a  field   of   nonzero
characteristics are called modular representations.
There exists a wide literature devoted to such representations;
detailed references can be found for example in Ref.
\cite{FrPa} (see also Ref. \cite{lev4}). In particular, it has
been shown by Zassenhaus \cite{Zass} that all modular irreducible
representations (IRs) are finite-dimensional and many papers have
dealt with the maximum dimension of such representations.
For classical Lie algebras this has been done in Refs.
\cite{BN,Zass,Rud,Rud1,Kac}. The complete classification of IRs
has been obtained only for $A_1$ algebras \cite{Rud}; 
important results
for $A_2$ and $B_2$ algebras are obtained in Refs. 
\cite{Braden,Jan};
results for $A_n$ algebras can be found in Refs. 
\cite{Jan,Pan,Dip},
and results obtained for some other algebras can be 
found in Ref.
\cite{Jan}. At the same time, it is worth noting that usually 
mathematicians consider only representations
over an algebraically closed field. 

From  the previous, it is natural to expect that the
correspondence  between
ordinary and modular representations of two Lie algebras over $R$ 
and $F_p$, respectively, can be obtained if the structure 
constants of the Lie algebra
over $F_p$ - $c_{kl}^j$, and the structure constants of the Lie
algebra over $R$ - ${\tilde c}_{kl}^j$, are 
connected by the relations
$f(c_{kl}^j)={\tilde c}_{kl}^j$ 
(the Chevalley basis \cite{Chev}), and
all the $c_{kl}^j$ belong to $U_0$. In Refs. \cite{lev4,lev3} 
there have been
considered modular analogs of IRs
of su(2), sp(2), so(2,3), so(1,4) algebras and the osp(1,4)
superalgebra. Also modular representations describing strings
have been briefly mentioned. In all these cases the quantities 
${\tilde c}_{kl}^j$ take only the values $0,\pm 1,\pm 2$ and there 
have been shown that the correspondence under consideration does 
have a place.

One might wonder how continuous transformations (e.g. 
time evolution or rotations) can be described in the framework 
of the GFQT. A general remark is that if theory ${\cal B}$ is a
generalization of theory ${\cal A}$ then the relation between
the theories is not always straightforward. For example, 
quantum mechanics is a generalization of classical
mechanics, but in quantum mechanics the experiment outcome
cannot be predicted unambiguously etc. As
noted in Sect. \ref{S2}, even in the framework of standard
quantum theory, the time evolution is well-defined only if $t$ is
a good approximate parameter. Suppose that this is the case, and
the Hamiltonian $H_1$ in standard theory is a good approximation
for the Hamiltonian $H$ in the GFQT. Then one might think that
$exp(-iH_1t)$ is a good approximation for $exp(-iHt)$. However,
such a straightforward conclusion is not correct for the following
reasons. First, there can be no continuous parameters in the GFQT.
Second, even if $t$ is somehow discretized, the notion of
$exp$ in the GFQT is not clear since in standard mathematics $exp$
is defined by an infinite Taylor series containing infinitely
small quantities. Therefore a direct correspondence between the
standard quantum theory and the GFQT in the sense described in 
this section exists only between Lie algebras but not Lie groups.
However, at some conditions (e.g. when $t$ is discretized and
only a finite number of the Taylor series for $exp$ is important)
one can define $exp(-iHt)$ in the GFQT. Analogously one can
define rotations, Lorentz boosts etc.

\section{Modular analogs of single-particle and two-particle
representations of the so(1,4) algebra}
\label{SG3}

As noted in Sect. \ref{S2}, in the framework of our approach
the dS invariance on quantum level implies 
{\it by definition} that the
operators describing the system satisfy the commutation
relations (\ref{2}). Therefore in the GFQT one could postulate 
that the same relations should be valid if the operators act in
a space over $F_{p^2}$. However, in that case the following question
arises. The relations in Eq. (\ref{2}) contain $i$ explicitly
while, as noted in Sect. \ref{SG1}, the field $F_{p^2}$ contains
the element which can be identified with $i$ only if 
$p=3\,\,(mod\,\,4)$. A possible solution is as follows. As noted
in Sect. \ref{S8}, in terms of the operators 
$({\bf J}'{\bf J}"R_{ij})$ the commutation 
relation (\ref{2}) can be 
written in the form of Eqs. (\ref{14}-\ref{16}) which do not
contain $i$ explicitly. One might think that Eqs. 
(\ref{14}-\ref{16}) are rather chaotic, but in fact they are
very natural in the Weyl basis of the so(1,4) algebra. We 
therefore can postulate
that the dS invariance in the GFQT implies 
{\it by definition} that
Eqs. (\ref{14}-\ref{16}) are satisfied.

The next step is the construction of modular analogs of UIRs of
the so(1,4) algebra. One can define the operators
$A^{++}$, $A^{+-}$, $A^{-+}$ and $A^{--}$ by Eq. (\ref{18})
which now should treated in spaces over $F_{p^2}$.
By analogy with the construction in Sect.
\ref{S8} one can require that in a space over 
$F_{p^2}$ there
exists a vector $e_0$ satisfying Eq. (\ref{26}) but $w$ and $s$
should now be treated as elements of $F_p$. Then, proceeding as
in Sect. \ref{S8}, one can derive Eq. (\ref{28}).

As noted in Sect. \ref{S8}, in the standard theory Eq. 
(\ref{28}) would imply that the quantum number $n$ can take
the values $n=0,1,...$ and therefore the representation is
infinite dimensional. However, in the modular case $n$ can
take only the values $0,1,...n_{max}$ where the maximum 
value of $n$, $n_{max}$ can be found as follows. By definition
of the operator $A^{++}$, $A^{++}e^{nr}=0$ if $n=n_{max}$.
This relation should not contradict Eq. (\ref{28}) if 
$n=n_{max}$. Therefore $n_{max}$ should satisfy the condition
\begin{equation}
(n_{max}+1)(n_{max}+s+2)[w+(2n_{max}+s+3)^2]=0
\label{G13}
\end{equation} 

As follows from Eq. (\ref{G13}), $n_{max}$ cannot be greater
than $p-s-2$ and therefore the irreducible representation 
is necessarily finite dimensional in agreement with the
Zassenhaus theorem \cite{Zass}. In view of the correspondence
principle discussed in the preceding section, it is natural to
require that $s\ll p$ and then $n_{max}$ is of order $p$. 
However, there also exists a possibility that $n_{max}$ is
much less than $p$ if $w+(2n_{max}+s+3)^2=0$ in $F_p$. In the standard
theory $w=\mu^2$ where $\mu$ is the particle dS mass. In the
GFQT the element $w\in F_p$ is not necessarily a square in $F_p$
but suppose that this is the case. As noted in Sect. \ref{SG1},
the prime number $p\neq 2$ can be represented as a sum of two squares
if and only if $p=1\,\,(mod\,\,4)$. Therefore, if $w$ is a square
in $F_p$ and $p=3\,\,(mod\,\,4)$ then the equality 
$w+(2n_{max}+s+3)^2=0$ in $F_p$ is impossible and $n_{max}=p-2-s$.
However, if $w$ is a square in $F_p$ and $p=1\,\,(mod\,\,4)$ then the 
equality $w+(2n_{max}+s+3)^2=0$ in $F_p$ is possible and typically
$n_{max}$ is less than $p-2-s$. There exist scenarios (see Ref.
\cite{lev3}) when $n_{max}$ is of order $p^{1/2}$, but in any
case, if $p$ is large, the dimension of the irreducible 
representation is quite sufficient to ensure the correspondence
principle. 

The reader can notice that if the dS invariance is postulated in 
the form of Eqs. (\ref{14}-\ref{16}) then the consideration in
Sect. \ref{S8} does not involve $i$ at all: we only assume that,
according to the principles of quantum theory, the operators
should act in complex Hilbert spaces. To ensure the correspondence
with standard quantum theory, we assume that in GFQT the
operators act in spaces over $F_{p^2}$. However, if GFQT is
treated as a generalization of standard quantum theory then
there should be deeper reasons for that. If we were able to find 
reasons why the operators in GFQT should act in spaces $F_{p^2}$
this could explain why standard quantum theory is based on the
field of complex numbers. This problem will be considered
elsewhere.

By analogy with the consideration in Sect. \ref{S11}, one can
consider a system of two spinless particles in the GFQT. In
particular, the element $\Phi_n$ can be again defined by
Eq. (\ref{60}) assuming that the operators act in the tensor
product of two modular IRs. Then the
$(\Phi_n,\Phi_n)$ is formally given by Eq. (\ref{61})
where the r.h.s. should be understood as an element of $F_p$.
Analogously one can define the operator $W$ such that its action
on $\Phi_n$ is given by Eq. (\ref{64}) and its matrix elements
are given by Eq. (\ref{65}).

In contrast with the situation in the standard theory, where the
space $H_s$ is infinite dimensional, in the GFQT it is always
finite dimensional. This is already clear from the fact (see
Eq. (\ref{60})) that $\Phi_n$ is a linear combination of the
products of $e_{\alpha\beta}^{(1)n+j}$ and 
$e_{\alpha\beta}^{(2)n+j}$. Therefore if $N_{1max}$ is the
maximum value of $n$ in $e_{\alpha\beta}^{(1)n}$ and $N_{2max}$
is the maximum value of $n$ in $e_{\alpha\beta}^{(2)n}$ then
the maximum value of $n$ for $\Phi_n$ is   
\begin{equation}
N_{max} = min(N_{1max}-\frac{s}{2},N_{2max}-\frac{s}{2})
\label{G14}
\end{equation}
A detailed discussion of the spectrum of $W$ in the modular case
is given in Ref. \cite{lev3}. 

As noted in Sect. \ref{SG1}, 
division in $F_p$ in general considerably differs from 
that in the field of rational numbers. Therefore the important
problem arises whether the presence of division in 
Eq. (\ref{65}) is compatible with the correspondence between
the GFQT and the standard theory for the mean value of the
operator $W$.

Let us consider the notion of the mean value in the GFQT.
In the standard theory the wave function $\Psi$ of any state can be
always normalized to one, and in that case the mean value of the
operator $A$ in the state $\Psi$ is defined as 
${\bar A}=(\Psi,A\Psi)$. As noted in the preceding section,
normalization to one is assumed only for convenience and does not
have any fundamental meaning. If the state $\Psi$ is not normalized
to one then the mean value of the operator $A$ in the standard
theory is defined as
\begin{equation} 
{\bar A}=\frac{(\Psi,A\Psi)}{(\Psi,\Psi)}
\label{G15}
\end{equation}
In the GFQT it is also possible to normalize states to one,
but, since $1/n$ is typically a big number in $F_p$, such a
normalization is meaningless. However, as discussed in the
preceding section, if $|f(\Psi,A\Psi)|\ll p$, 
$|f(\Psi,\Psi)|\ll p$ and $f(\Psi,\Psi) > 0$ then one can
define probabilities as in the standard theory. In that case
the numerator and denominator in Eq. (\ref{G15}) can be
treated as usual integers and their ratio should be calculated
in a standard way (i.e. as in the field of rational numbers).

Suppose that $\Psi$ is a state in the
modular analog of $H_s$ and
\begin{equation}
\Psi=\sum_{n=0}^{N_{max}} c_n\Phi_n
\label{G16}
\end{equation} 
where the $c_n$ are elements of $F_{p^2}$. Suppose also 
that the 
quantities $c_n$ are different from zero only for $n\ll p$ and 
for such values of $n$ the conditions $|f(c_n{\bar c}_n)|\ll p$
and $f(c_n{\bar c}_n) > 0$ are satisfied. Then there exists the
correspondence between the quantities $c_n$ in the standard
theory and GFQT. If $|f(w_j)|\ll p$, $f(w_j)> 0$ $(j=1,2)$ and
$n\ll p$ then the correspondence between the quantities
$(\Phi_n,\Phi_n)$ exists as a consequence of Eq. (\ref{61})
and the fact that
\begin{equation}
\frac{(n+s+1)!}{(n+1+s/2)n!}=(n+1)\cdots (n+s/2)(n+s/2+2)
\cdots (n+s+1)
\label{G17}
\end{equation} 
i.e. the denominator in Eq. (\ref{61}) is cancelled out by the
numerator. Since
\begin{equation}
(\Psi,\Psi)=\sum_{n=0}^{N_{max}}c_n{\bar c}_n(\Phi_n,\Phi_n)
\label{G18}
\end{equation}
the correspondence for
$(\Psi,\Psi)$ in the standard theory and GFQT exists too. 
Note that if the
coefficients $c_n$ are such as discussed above then actually
the upper limit in the sum is much less than $N_{max}$.

\begin{sloppypar}
Conside now the correspondence between the
quantities $(\Psi,W\Psi)$ in the standard theory and GFQT. 
As follows
from Eqs. (\ref{61}), (\ref{64}), (\ref{65}) and (\ref{G16})
\begin{eqnarray}
&&(\Psi,W\Psi)=\sum_{n=0}^{N_{max}}W_{nn}(\Phi_n,
\Phi_n)c_n{\bar c}_n+\nonumber\\ 
&&\sum_{n=0}^{N_{max}}\frac{n+2+s}{n+2+s/2}(\Phi_n,
\Phi_n)[w_1+(2n+s+3)^2]\nonumber\\
&&[w_1+(2n+s+3)^2](c_{n+1}{\bar c}_n+c_n{\bar c}_{n+1})
\label{G19}
\end{eqnarray}
One can notice again that actually 
the upper limit is much less than $N_{max}$.
Since there is no division in $W_{nn}$ (see Eq. (\ref{65})),
no problem arises with the correspondence for the first
sum in the r.h.s. of this expression. Since $(\Phi_n,\Phi_n)$
contains $(n+1+s)!/[(n+1+s/2)n!]$, the second sum in 
Eq. (\ref{G19}) contains
\begin{equation}
\frac{(n+2+s)!}{(n+1+s/2)(n+2+s/2)n!}=\frac{(n+1)\cdots (n+2+s)}
{(n+1+s/2)(n+2+s/2)}
\label{G20}
\end{equation}
It is obvious that the denominator is again cancelled out by the
numerator and therefore there exists the correspondence for the
quantity $(\Psi,W\Psi)$. 
\end{sloppypar}

Our final conclusion in this section is that the presence of
division in nondiagonal matrix elements of $W$ is
not an obstacle for the correspondence of the mean 
values of this operator in the standard theory and GFQT.

\section{Discussion}

In the present paper we have investigated 
several aspects of the property
that, in contrast with the Poincare and AdS invariant
theories, the mass operator in the dS invariant theory is 
not bounded below by the value of $m_1+m_2$. 

We treat the dS invariance on quantum level as the
requirement that the operators describing the system
should satisfy the commutation relations (\ref{2}). In
contrast with the approach based on gauge theories in
curved spacetime, we do not require the existence of any
classical background (i.e. the dS space). 
   
In Chap. \ref{Ch2} we
have considered the results which can be obtained with 
rather simple calculations. They are based on the form of 
UIRs of the dS algebra resembling the main features of 
UIRs of the
Poincare algebra. It has been shown that the unusual
features of the dS invariant theory are manifested already
in first order in $1/R$, where $R$ is the dS radius.
Although the phenomenon of the dS antigravity is well
known, we pay attention to the fact
that the interaction caused by the dS antigravity has
not been introduced as an interaction operator. On the
level of dS invariance the particles are treated as free 
since they are described by the tensor product of UIRs; 
the interaction arises
only when one treats the particles in Poincare (or Galilei)
invariant terms. This observation poses the problem 
whether all interactions in nature can be obtained as
a result of transition from a higher symmetry to a
lower one. Analogous ideas have been proposed in
Kaluza-Klein theories, superstring theories etc.
In our approach no extra dimensions or new objects are
introduced.

As discussed in Sects. \ref{S6} and \ref{classical}, the 
fact that
the free two-particle mass operator in the dS invariant
theory is not bounded below by the value of $m_1+m_2$
does not mean that the theory is unphysical. Moreover,
this fact opens new possibilities for understanding the
physical meaning of interactions. As argued in Sect.
\ref{S7}, one might expect that for any reasonable
interaction, the free and interacting mass operators are
unitarily equivalent. 

In Chap. \ref{Ch3} we develop an approach based on the 
su(2)$\times$su(2) basis,
and the results are used in Chap. \ref{Ch4}. In particular,
we derive an important Eq. (\ref{84}) for the mean
value of the two-body mass operator. Then we show that
the standard relativistic result (\ref{85}) is a
consequence of Eq. (\ref{84}) in a special case when
the decomposition coefficients $c_n$ and $c_{n+1}$ are
approximately the same. 

If the phases of $c_n$ and $c_{n+1}$ are 
different, then, as follows from Eq. (\ref{84}), 
there exists a correction to the standard relativistic
result, which is always negative and proportional to
the particle masses in the nonrelativistic approximation.
In particular, the correction corresponding to the
standard Newtonian potential $-Gm_1m_2/r$ is obtained
in the case where the phase difference between 
$c_n$ and $c_{n+1}$ is equal to 
$\delta=[2G(m_1+m_2)/r]^{1/2}$.
Since the quantity $2Gm$ is the gravitational radius for
the particle with the mass $m$, the condition $\delta\ll 1$
is exactly the well known requirement that Newtonian gravity
is the good first approximation to General Relativity if
the particles are nonrelativistic and their gravitational
radii are much less than the distance between them. 
As shown in Sect. \ref{phase}, analogous results take place
not only in the nonrelativistic but also in post-Newtonian
approximation.

As we have already mentioned, a very interesting possibility
is that gravity can be explained without introducing any
interaction at all. Our hypothesis is that this can be
implemented in quantum theory over a Galois field (GFQT). 
One of the reasons is as follows. Since the spectrum of
the two-body mass operator in dS invariant theories is
not bounded below by $m_1+m_2$, the nonrelativistic
mass operator is not positive definite. Consider a
quasiclassical two-body system with the relative momentum
$q$ and relative distance $r$. If the rules of calculating 
probabilities are not exactly the same as in the standard 
theory then the most probable value of the 
two-body mass might be $q^2/2m_{12}-Gm_1m_2/r$ rather
than $q^2/2m_{12}$. 

In Chap. \ref{Ch5} the basics of the GFQT is 
described. We argue that it is not only simple and 
elegant but also a more natural quantum theory than the 
standard one. As noted in Sect. \ref{SG1}, the GFQT
can be treated as a version of the Heisenberg matrix
formulation of quantum theory when the field of
complex numbers is replaced by a Galois field. In that
case the theory does not contain actual infinity and
therefore the problem of divergencies does not exist
in principle. 

A well known point of view on the hierarchy of
physical theories is that (see e.g. Ref. \cite{FJD2}
and references therein)
any quantum theory with a lower symmetry can be obtained
from some theory with a higher symmetry by using the
contraction procedure. For example, the nonrelativistic
quantum mechanics can be obtained from relativistic one
by means of contraction $c\rightarrow \infty$. 
In turn, the latter can be obtained from
dS or AdS invariant theories by means of contraction 
$R\rightarrow \infty$. Since the dS and AdS groups are 
semisimple, the
dS and AdS invariant theories cannot be obtained from
another theory by means of any contraction. However, this
statement is based on the assumption that the number field
used in all the theories is the same. As noted in
Chap. \ref{Ch5}, the transition from a theory based on $F_p$ to
one based on the ring of integers can be treated as a
contraction $p\rightarrow \infty$. 

One might wonder whether the GFQT could be treated as
a version of quantum theory with a fundamental
length. For the first time the notion of 
fundamental length has been probably discussed by
Snyder \cite{Snyder}. He has considered a possibility
that different coordinate operators do not commute
with each other and their commutator is proportional
to a quantity with the dimension $(length)$. At
present this idea has become popular in view of
quantum theories on noncommutative spaces (see e.g.
Ref. \cite{noncomm} and references therein).
In the GFQT all physical quantities are
dimensionless and take values in a Galois field.
Neverthless, on a qualitative level the GFQT can
be thought to be a theory with the fundamental
length in the following sense. The maximum value
of the angular momentum in the GFQT cannot exceed
the characteristic of the Galois field $p$. Therefore
the Poincare momentum cannot exceed $p/R$. This
can be qualitatively interpreted in such a way that
the fundamental length in the GFQT is of order $R/p$.

As noted in Chap. \ref{Ch2}, in the standard approach there
exist infinitely many forms of UIRs and they are
related to each by unitary transformations (for example,
the forms (\ref{9}) and (\ref{12}) are related to
each other by the unitary transformation (\ref{13})
containing a large exponent index). Since the
interaction depends on the form, the question arises,
what form is physically
preferable. In the framework of the GFQT the answer
is clear: the form (\ref{12}) is preferable since it
can be reformulated in terms of modular representations. 

As shown in Sect. \ref{SG3}, starting from modular
irreducible representations one can extend the results
of the standard theory and calculate the mean value of
the two-body mass operator in the GFQT. We will now
discuss our hypothesis that in such a way it will be
possible to find a natural explanation of gravity.

The problem of cosmological constant is one of the most
outstanding in modern quantum theory and there exists a wide
literature on this problem (see e.g. Ref. 
\cite{Weinberg,Einst-dS}
and references therein). The explanation of the problem
"for pedestrians" is rather simple. Naive calculations
of the cosmological constant $\Lambda$ in quantum field 
theory produce
anomalously large values and this could be expected
from the beginning. Indeed, the Planck length is of order
$l_P=10^{-33}cm$ and therefore if $\Lambda$ is of order unity
in the Planck units then it is of of order $10^{66}/cm^2$
in the units $\hbar/2 = c =1$. Meanwhile, since 
$\Lambda =3/R^2$, for realistic values of $R$ it is much
smaller (for example, if $R$ is 20 billion light years
(in qualitative agreement with the data of Ref. 
\cite{Perlmutter})  then
$\Lambda$ is of order $10^{-56}/cm^2$ and the discrepancy
is 123 orders of magnitude).

As noted in Sect. \ref{S2}, in our approach we postulate
that on quantum level the dS invariance implies that
the expressions in Eq. (\ref{2}) are valid. They do not
contain the cosmological constant at all and therefore
in such a formulation the problem of cosmological 
constant does not exist. This is clear from the fact that
all the operators describing the dS algebra are angular
momenta and therefore in units $\hbar/2 = c =1$ there are
no quantities with the dimensions $(length)^n$ 
$(n=\pm 1, \pm 2,...)$. Such quantities arise only if one
wishes to reformulate the results in terms of Poincare
invariant theories or classical dS space. In other words, 
on quantum level any dS invariant theory is described 
only in terms of angular momentum and/or angular 
variables (a discussion on whether 
angular or radial variables are more fundamental can
be found, for example, in Ref. \cite{angle}).

In units $\hbar/2 = c=1$ the gravitational constant $G$ is of
order $10^{-66}cm^2$ since $G=l_P^2$. Let us rewrite the
Newtonian potential in dS terms. Since the dS energy is
$M_{04}=2RE$ and the dS masses are related to Poincare
ones as $\mu = 2Rm$ (see Sect. \ref{S5}) then the Newtonian
potential $-Gm_1m_2/r$ in dS terms is 
$-{\tilde G}\mu_1\mu_2/\varphi$ where the angular variable
$\varphi$ has the meaning of $r/(2R)$ and 
${\tilde G}=G/(4R^2)$.

If one accepts that dS invariance is more fundamental
than Poincare one then it is ${\tilde G}$ which should be
treated as a "true" gravitational constant. 
Note that ${\tilde G}$ is proportional to $G\Lambda$. 
This quantity is dimensionless and extremely small 
(for example, if $R$ is 20 
billion light years then ${\tilde G}$ is of order $10^{-123}$).
Therefore the fact that $G$ cannot be neglected in our world
can be explained as a consequence of the circumstance that
$G=4R^2{\tilde G}$, i.e. a very small quantity ${\tilde G}$
is multiplied by a very large one $4R^2$. 

If one accepts that $G$ and $\Lambda$ are no longer
fundamental then there is no reason to believe that they
are really constants. Since $G$ is proportional to 
$R^2$ and $\Lambda$ is proportional to $1/R^2$, it
seems reasonable to suggest that at present $G$ is
increasing and $\Lambda$ is decreasing (in contrast
with Dirac's hypothesis \cite{Dirhyp}). 

Our hypothesis
is that ${\tilde G}$ is different from zero as a consequence
of the fact that quantum theory is based on a Galois field
rather than the field of complex numbers. In other words,
${\tilde G}=1/p_0$ where the large number $p_0$ is somehow
related to the characteristic of the Galois field $p$.
In that case if formally $p\rightarrow \infty$ then 
${\tilde G}\rightarrow 0$. Therefore, the transition 
${\tilde G}\rightarrow 0$ in our hypothesis
has the meaning of contraction from the GFQT to the 
standard quantum theory.

To verify this hypothesis, one has to calculate the mean value
of the two-body mass operator in the framework of the GFQT.
As shown in Sect. \ref{SG3}, such a calculation is in
principle possible since the correspondence between the
standard theory and the GFQT is not broken as a consequence
of the fact that nondiagonal matrix elements of the $W$
operator contain division in $F_p$. The main technical 
problems are that quasiclassical states should be 
constructed in the framework of the GFQT,
and all the calculations should be performed
exactly (since there is no GFQT analog of the stationary
phase method).  

{\it Acknowledgements: } The author is grateful to 
B. Hikin, V.A. Karmanov, M.B. Mensky, M.A. Olshanetsky, 
M. Partensky and M. Saniga for fruitful discussions.

\end{document}